\documentstyle{mn}
\input {psfig.sty}
\def \Mpc {~h^{-1}~{\rm Mpc} }
\def \Om {\Omega_0}

\def \lo {\lambda_0}

\def \bj {b_{\rm J}}
\def \qso {_{\rm Q}}
\def \gal {_{\rm gal}}
\def \xibar {\bar{\xi}}
\def \max {_{\rm max}}
\def \gsim { \lower .75ex \hbox{$\sim$} \llap{\raise .27ex \hbox{$>$}} }
\def \lsim { \lower .75ex \hbox{$\sim$} \llap{\raise .27ex \hbox{$<$}} }
\def \deg {^{\circ}}
\def \cobe {{\it COBE }}
\def \deltac {\delta_{\rm c}}
\def \mmin {M_{\rm min}}
\def \msun {M_{\odot}}

\title[2dF QSO Survey: Clustering]
      {The 2dF QSO Redshift Survey - II.  Structure and evolution at
      high redshift}
\author[S.M. Croom et al.]
       {Scott M. Croom$^{1,2}$\thanks{scroom@aaoepp.aao.gov.au},
      T. Shanks$^3$, B.J. Boyle$^2$, 
      R.J. Smith$^4$, L. Miller$^5$, \\
{\LARGE N.S. Loaring$^5$ \& F. Hoyle$^3$}\\
$^1$Imperial College of Science, Technology and
      Medicine, Blackett Laboratory, Prince Consort Road, London, SW7
      2BW, UK.\\
$^2$Anglo-Australian Observatory, PO Box 296, Epping, NSW 2121,
      Australia.\\
$^3$Physics Department, University of Durham, South Road, Durham, DH1 3LE,
UK.\\
$^4$Research School of Astronomy and Astrophysics, ANU, Private bag,
      Weston Creek P.O., ACT 2611, Australia.\\
$^5$Department of Physics, Oxford University, Keble Road, Oxford, OX1
      3RH, UK.}
\begin{document}

\maketitle

\begin{abstract}
In this paper we present a clustering analysis of QSOs over the
redshift range $z=0.3-2.9$.  We use a sample of 10558 QSOs taken from
preliminary data release catalogue of the 2dF QSO Redshift Survey
(2QZ).  The two-point redshift-space correlation function of QSOs,
$\xi\qso(s)$, is shown to follow a power law on scales
$s\simeq1-35\Mpc$.  Fitting a power law of the  form
$\xi\qso(s)=(s/s_0)^{-\gamma}$ to the QSO clustering averaged over the
redshift interval $0.3<z\leq2.9$ we find
$s_0=3.99^{+0.28}_{-0.34}\Mpc$ and $\gamma=1.58^{+0.10}_{-0.09}$ for
an Einstein-de Sitter cosmology.  The effect of a significant
cosmological constant, $\lo$, is to increase the separation of QSOs,
so that with $\Om=0.3$, $\lo=0.7$ the power law extends to
$\simeq60\Mpc$ and the best fit is $s_0=5.69^{+0.42}_{-0.50}\Mpc$ and
$\gamma=1.56^{+0.10}_{-0.09}$. These values, measured at a mean
redshift of $\bar{z}=1.49$, are comparable to the clustering of local
optically selected galaxies.  We compare the clustering of 2QZ QSOs to
generic CDM models with shape parameter $\Gamma_{\rm eff}$.  Standard
CDM with $\Gamma_{\rm eff}=0.5$ is ruled out in both Einstein-de
Sitter and cosmological constant dominated cosmologies, where
$\Gamma_{\rm eff}\simeq0.2-0.4$ and $\Gamma_{\rm eff}\simeq0.1-0.2$
respectively are the allowable ranges.

We measure the evolution of QSO clustering as a function of redshift.
For $\Om=1$ and $\lo=0$ there is no significant evolution in comoving
coordinates over the redshift range of the 2QZ.  QSOs thus have
similar clustering properties to local galaxies at all redshifts we
sample.  In the case of  $\Om=0.3$ and $\lo=0.7$ QSO clustering shows
a marginal increase at high redshift, $s_0$ being a factor of
$\sim1.4$ higher at $z\simeq2.4$ than at $z\simeq0.7$.  Although the
clustering of QSOs is measured on large scales where linear theory
should apply, the evolution of QSO clustering does not follow the
linear theory predictions for growth via gravitational instability
(rejected at the $>99$ per cent confidence level).  A redshift
dependent bias is required to reconcile QSO clustering observations
with theory.  A simple biasing model, in which QSOs have
cosmologically long lifetimes (or alternatively form in peaks above a
constant threshold in the density field) is acceptable in an $\Om=1$
cosmology, but is only marginally acceptable if $\Om=0.3$ and
$\lo=0.7$. Biasing models in which QSOs are assumed to form over a
range in redshift, based on the Press-Schechter formalism, are
consistent with QSO clustering evolution for a minimum halo mass of
$\sim10^{12}\msun$ and $\sim10^{13}\msun$ in an Einstein-de Sitter and
cosmological constant dominated universe, respectively.  However,
until an accurate, physically motivated, model of QSO formation and
evolution is developed, we should be cautious in interpreting the fits
to these biasing models.
\end{abstract}


\section{Introduction}

The 2dF QSO Redshift Survey (2QZ) aims to compile a homogeneous
catalogue of $\sim25000$ QSOs using the Anglo-Australian Telescope
(AAT) 2-degree Field facility (2dF; Taylor, Cannon \& Watson 1997).
This catalogue will constitute a factor of $\gsim50$ increase in
numbers to a equivalent flux limit over previous data sets (e.g. Boyle
et al. 1990).  The main science goal of the 2QZ is to use QSOs  to
probe the large-scale structure of the Universe over a range of scales
from 1 to $1000\Mpc$, and in the redshift interval, $0.3\lsim z\lsim2.9$.

Clustering of QSOs at small to intermediate scales ($1-50\Mpc$)
supplies a wealth of information on large-scale structure.  QSOs still
give us the only method of directly determining the 3-dimensional
clustering of high redshift objects within a large enough volume for
it to be truly representative.  When complete, the 2QZ
will sample a volume of $1.5\times10^9$h$^{-3}$Mpc$^3$ (for $\Om=1$),
an order of magnitude larger than current galaxy redshift surveys
(e.g. the 2dF Galaxy Redshift Survey; Colless 1999).  This large
volume also allows us to probe the scales where linear evolution
occurs, simplifying comparisons with theory.

The shape and amplitude of the two-point auto-correlation function,
$\xi(r)$, is determined by two factors.  The first is the distribution
of matter fluctuations in the Universe.  This depends on fundamental
physics, such as the growth of structure via gravitational instability
and the initial spectrum of fluctuations.  The second factor concerns
the complex and generally non-linear physics which occurs during
galaxy and QSO formation.  The difference between the matter and
galaxy or QSO distributions is commonly called bias, $b(r,z)$, such that
\begin{equation}
\xi\qso(r,z)=b^2(r,z)\xi_{\rho}(r,z),
\end{equation}
where $\xi\qso(r,z)$ and $\xi_{\rho}(r,z)$ are the two-point
correlation functions of QSOs and the density field respectively.
Both are functions of scale, $r$, and redshift, $z$.  Often, a {\it
linear bias} is assumed, which has no scale dependence, and it appears
likely that for any local process of galaxy formation $b$ should tend
to a constant value on scales where the density perturbations are
linear (e.g. Mann, Peacock \& Heavens 1998; Peacock 1997).  We will
assume a linear bias throughout this paper.

The first attempt to measure the clustering of QSOs was made by Osmer
(1981).  Shaver (1984) was the first to detect QSO clustering on small
scales, although in an inhomogeneous sample. Shanks et al. (1987) made
the first detection of clustering at $\lsim10\Mpc$ in a complete and
uniformly selected sample; part of the Durham/AAT UVX survey
\cite{bfsp90}.  A number of authors have used this and other QSO
samples to measure clustering.  They all reach generally the same
conclusions that clustering is detected at the $\sim3-4\sigma$ level
and is approximately consistent with a clustering scale length
$r_0\sim6\Mpc$, similar to local galaxy clustering, at a mean redshift
of $z\sim1.4$ \cite{is88,ac92,mf93,sb94,cs96}.  There has been significant
disagreement over the redshift evolution of QSO clustering including
claims for a decrease in the QSO correlation length ($r_0$) with
redshift \cite{is88}, an increase in $r_0$ with redshift (La Franca,
Andreani \& Cristiani 1998; hereforth LAC98) and no change with
redshift (Croom \& Shanks 1996; hereforth CS96).

The measurement of galaxy clustering at high redshift has also taken
dramatic steps forward in recent years.  A number of surveys have made
measurements of the clustering strength of galaxies up to $z\sim1$.
These samples typically contain a few hundred to a thousand galaxies
over relatively small areas.  The Canada-France Redshift Survey (CFRS)
shows a significant decrease in clustering amplitude on scales
$<2\Mpc$ \cite{cfrs96}.  However, larger samples, such as the CNOC-2
survey \cite{cnoc2} show much slower evolution, with a gradual
decrease of clustering with redshift:
$r_0(z)\propto(1+z)^{-0.3\pm0.2}$.  Deep wide-field ($\sim$ few
degrees) imaging surveys used to measure the angular correlation
function of galaxies also suggest higher clustering amplitudes than
found in the CFRS \cite{plso98}.  The differences found between these
samples is partly due to the different selection methods
(e.g. magnitude limits; photometric bands) used.  It is likely that
different galaxy types cluster differently, e.g. optically vs.
infrared selected galaxies (Peacock 1997).  Clustering is also likely
to be a  function of galaxy luminosity.  It is possible that this is
the case for QSOs, although we leave the discussion of luminosity
dependent clustering of QSOs to a future paper.  We note, however,
that due to  the stong luminosity evolution of QSOs (e.g. Boyle et
al. 2000) an apparent magnitude limited survey of QSOs samples
approximately the same part of the luminosity function at all redshifts up
to $z\sim2$.  A second affect responsible for the difference in galaxy
clustering results is {\it cosmic variance} due to the small  volumes
and scales sampled, in particular by the CFRS.  A key element of the
2QZ is that it is large enough to minimize any effects of cosmic
variance on scales smaller than a few hundred $\Mpc$.  Although
studies of galaxy clustering have been typically limited to $z\lsim1$,
Steidel et al. (1998) have used galaxies detected by their Lyman-break
to derive the clustering properties of galaxies at $z\sim3$.  These
observations show that the clustering of $L\sim L^*$ galaxies at
$z\sim3$ is also similar to local galaxies on scales $\lsim10\Mpc$,
with $r_0\simeq4-6\Mpc$ depending on the assumed cosmology
\cite{adelberger98}.  

In this paper we look at QSO clustering in the 2QZ on scales from
$\sim1$ to $100\Mpc$.  We do not attempt to study larger scales because
of the current non-uniformity of the data set.  This will be reserved
for future work, on completion of the survey.  In Section
\ref{section_data} we describe the 2QZ data used and our methods of
analysis.  In Sections \ref{section_xir} and \ref{section_evol} we
present our clustering results and compare them to physical models.
Our conclusions are presented in Section \ref{section_conclusions}.

\section{Data and Techniques}\label{section_data}

\subsection{The 2dF QSO Redshift Survey}

\begin{figure*}
\centering
\centerline{\psfig{file=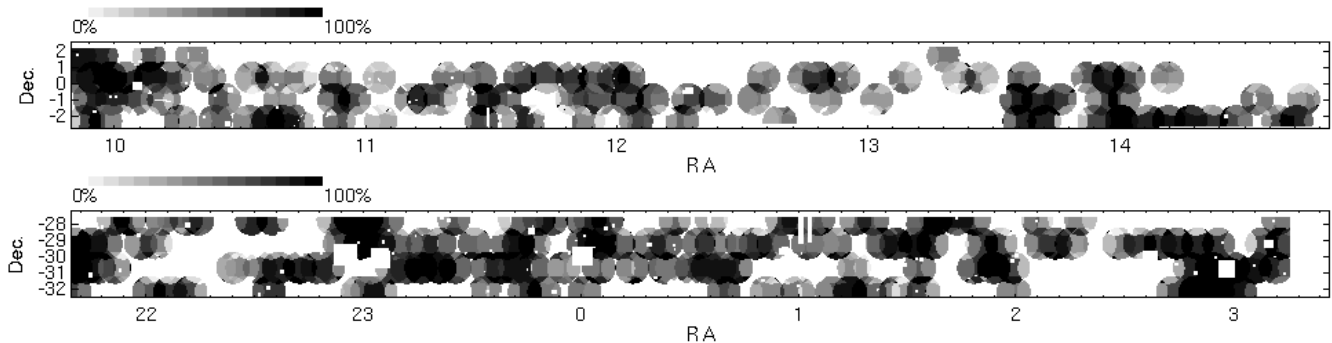,width=18.0cm,angle=270}}
\caption{The current fractional coverage in the NGC (top) and SGC
(bottom) strips of the 2QZ.  Each circular region corresponds to one
2dF pointing.  The coverage is the fraction of QSO candidates observed
in each region.  The small
rectangular holes correspond to regions containing bright stars and
plate defects.}
\label{coverage}
\end{figure*}

For the analysis in this paper we have used the first public release
catalogue of the 2QZ containing 10681 QSOs (the {\it 10k catalogue}).
This 10k catalogue contains the most spectroscopically complete fields
observed prior to November 2000 and will be released to the
astronomical community in the first half of 2001.  The sample contains
10558 QSOs in the redshift range $0.3<z\leq2.9$ which will be included
in our analysis below.

The identification of QSO candidates for the 2QZ was based on broad
band $ub_{\rm J}r$ colours from Automatic Plate Measuring (APM)
facility measurements of UK Schmidt Telescope (UKST)  photographic
plates.  The survey comprises 30 UKST fields, arranged in two
$75^{\circ}\times5^{\circ}$ declination strips centred in the South
Galactic Cap (SGC) at $\delta=-30^{\circ}$ and the North Galactic Cap
(NGC) at $\delta=0^{\circ}$ with RA ranges $\alpha=21^h40$ to $3^h15$
and $\alpha=9^h50$ to $14^h50$ respectively.  Each UKST field contains
independent CCD calibration \cite{phot1,phot2}.  The completed survey
will cover approximately 740 deg$^2$ (some areas having been removed
due to bright stars, plate defects etc).  Further details of the
photometric catalogue can be found in Croom (1997), Smith (1998) and
Smith et al. (2001).

Spectroscopic observations have been carried out using the 2dF
instrument at the AAT  in conjunction with the 2dF Galaxy Redshift
Survey \cite{2dfgrs}, as the 2QZ and galaxy survey areas cover the
same regions of sky.  Typically, 225 fibres are devoted to galaxies,
125 to QSO candidates and 25-30 to sky in each 2dF observation.
Spectroscopic data are reduced using the 2dF pipeline reduction system
\cite{2dfman}.  The identification of QSO spectra and redshift
estimation was carried out using the {\small AUTOZ} code written
specifically for this project (Miller et al. 2001 in preparation).
This program compares template spectra of QSOs, stars and galaxies to
the observed spectra.  Identifications are then confirmed by eye for
all spectra.  Spectroscopic completeness is typically $>80$ per cent
when observations are made in reasonable or good conditions.  In the
analysis below we use all objects which have been classified as class
1 QSOs (class 1 being the highest quality identification; objects
classified as class 2 IDs or ``QSO?'' were not included) and which
were observed in fields within the 10k catalogue (which is limited to
$\geq85$ per cent spectroscopic completeness).

\subsection{Correlation function estimates}\label{xiest}

As the QSO correlation function, $\xi\qso$ probes high redshifts and
large scales, the measured values are highly dependent on the assumed
cosmology.  We employ the method of Osmer (1981)\nocite{o81}, which
uses the coordinate transform in the Robertson-Walker metric
\cite{weinberg72} to determine the comoving separation of pairs of
QSOs.  We choose to calculate $\xi\qso$ for two representative
cosmological models; $\Om$,$\lo=$ $(1,0)$ and $(0.3,0.7)$, where $\Om$
and $\lo$ represent the conventional mass and vacuum energy
(cosmological constant) density contribution respectively, to the
total energy density of the Universe.  We will call these cosmological
models EdS (Einstein-de Sitter) and $\Lambda$ respectively. 

We have used the minimum variance estimator suggested by Landy \&
Szalay (1993)\nocite{ls93} to calculate $\xi(s)$, where $s$ is the
redshift-space separation of two QSOs.  This estimator is
\begin{equation}
\xi\qso(s)=\frac{QQ(s)-2QR(s)+RR(s)}{RR(s)},
\label{lseq}
\end{equation}
where $QQ$, $QR$ and $RR$ are the number of QSO-QSO, QSO-random and
random-random pairs counted at separation $s\pm\Delta s$.  We
bin our pairs such that $\log(\Delta s)=0.1$ or $0.2$.  The density of
random points used was $50$ times the density of QSOs.

The area of the survey is covered by a mosaic of 2dF pointings.  These
pointings overlap in order to obtain complete coverage in all areas,
including regions of high galaxy and QSO density.  As the survey is
not yet complete this means that certain areas within 2dF fields will
not have had all candidates observed, and therefore the observational
completeness of the sample varies strongly with angular position on
the sky.  This variation in observational completeness can clearly be
seen in Fig. \ref{coverage}.  Where a large number of 2dF pointing
overlap the coverage is $\sim100$ per cent, while in overlap regions
which have yet to be observed a second or third time the completeness
is significantly lower.  Particular care has been taken to construct
the random point distribution so as to take into account this angular
selection function.  In each region defined by the intersection of 2dF
fields we have counted the number of QSO candidates observed and
compared this to the total number to calculate the fractional
observational completeness.  We then weight the probability of a
random being placed in that region by this fractional completeness.
This corrects for the angular incompleteness due to overlapping 2dF
fields.  

\begin{figure}
\centering
\centerline{\psfig{file=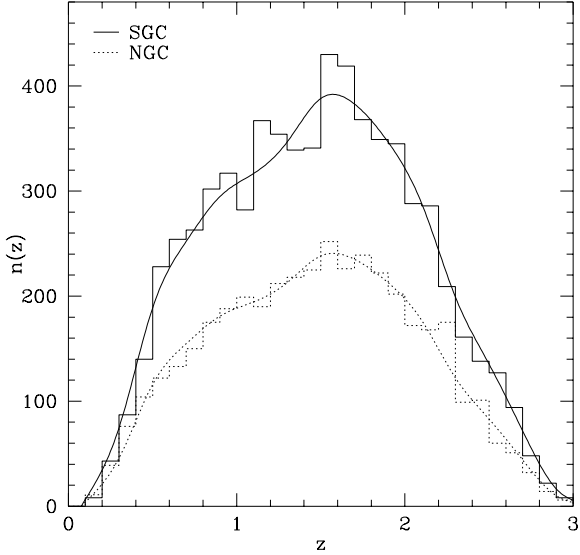,width=8.0cm}}
\caption{The redshift distribution of QSOs in the 10k catalogue for
the NGC (dotted histogram) and SGC (solid histogram) strips.  Also
shown is the fit used to generate random redshift distributions
(smooth curves), normalized to the number of QSOs in each slice.}
\label{nz}
\end{figure}

Our candidate density is not completely uniform over the length of the
strips, due to an increase in stellar contamination in areas closer to
the galactic plane.  Secondly, small residual calibration errors in
the relative magnitude zero points of the UKST plates could add
spurious structure on large scales.  Any possible offsets are being
corrected by calibration from further CCD photometry, however in this
paper we will correct for this effect by normalizing the number of
random points to the number of QSOs with spectroscopically determined
redshifts in each UKST field.  This correction will clearly remove
power on large scales, which is why we do not discuss structure on
scales larger than $\sim100\Mpc$ in this paper.  After constructing
the angular mask, we then assign the random points a random redshift,
taken from a spline fit to the binned ($\Delta z=0.2$) redshift
distribution of the full 2QZ sample (changing $\Delta z$ by a factor
of 2 makes no observable difference to $\xi\qso$).  The  redshift
distributions for the NGC and SGC data sets are shown in Fig \ref{nz}.
The fit used to generate the redshift distribution for the random
points is also shown (the smooth curves in Fig. \ref{nz}), in each
case normalized to the number of QSOs in each 2QZ strip.  The above
process for correcting observational completeness and calculating
$\xi\qso$ we call method 1.

\begin{figure}
\centering
\centerline{\psfig{file=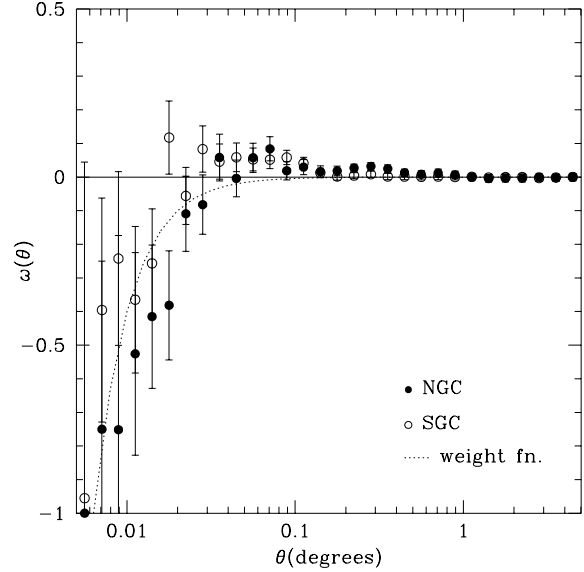,width=8.0cm}}
\caption{The angular correlation function for all currently observed
candidates in the 2QZ, split into the NGC (filled circles) and SGC
(open circles).  A deficit of pairs on small scales due to 2dF
positioning constraints can be seen.  The dotted line denotes
the weight function used to correct for the lack of close pairs.  The
uniformity at large scales, $>0.1\deg$, demonstrates the effectiveness
of our correction for the non-uniform field coverage.}
\label{candwtheta}
\end{figure}

We have tested the effectiveness of this process by making comparisons
to correlation functions derived using two other methods.  The first
(method 2) is to calculate the correlation function from regions of
the survey that have no overlapping fields still to be observed, that
is, they have $\sim100$ per cent observational coverage.  The number of QSOs
in these regions is significantly less that in the total sample,
reducing the signal-to-noise in $\xi\qso$.  The second comparison
method (method 3) is to allocate each random point an
($\alpha$,$\delta$) taken from the QSO catalogue, so that the random
distribution has {\it exactly} the same angular distribution as the
QSOs.  The redshifts of the random points are then allocated using the
spline fit discussed above.

\begin{figure*}
\centering
\centerline{\psfig{file=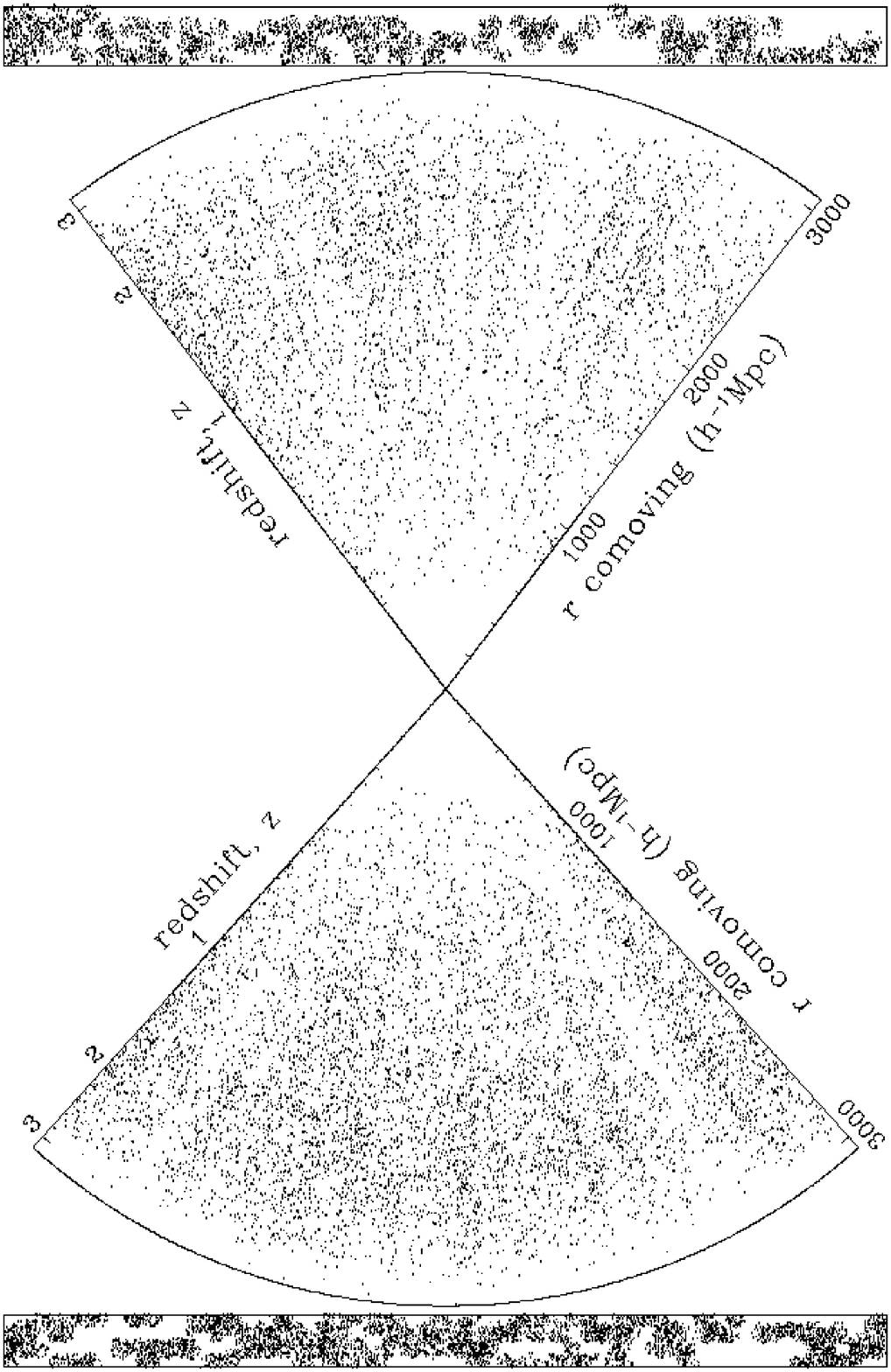,width=18.0cm,angle=270}}
\caption{The distribution of 2QZ QSOs in the 10k catalogue.  Note that
only objects at $0.3<z\leq2.9$ are used in our analysis. The SGC
strip is on the left, the NGC on the right.  The rectangular regions
show the distribution projected onto the sky in each strip.}
\label{wedge}
\end{figure*}

Two other observational biases could, in principle, affect our
measurements of $\xi\qso$.  The first is due to the fact that the 2dF
instrument cannot position two fibres closer than $\sim30''$.  We are
therefore currently biased against small angular separation QSO pairs
(this problem is being remedied by independent follow-up of close QSO
pairs).  We have measured the angular correlation function of observed
candidates, which shows this bias (see Fig. \ref{candwtheta}).
Measuring the extent of the anti-correlation in Fig. \ref{candwtheta}
allows us the correct for the close pairs bias.  The dotted line,
which traces the anti-correlation is
$\omega(\theta)=4.0\times10^{-5}\theta^{-2}$, and this can be used to
construct a function
\begin{equation}
W_{\rm cp}(\theta)=\frac{1}{1-4.0\times10^{-5}\theta^{-2}},
\end{equation}
which is the weight function for close pairs separated by $\theta$
degrees.  In practice this correction makes no difference to the
measured correlation function as almost all of the QSO pairs with
small angular separations have widely differing redshifts, and the
weighting of a small number of pairs has a negligible effect on large
scales.  

Extinction by galactic dust will also imprint a signal on the
angular distribution of the QSOs.  Primarily this changes the
effective magnitude limit in $\bj$ by $A_{\rm b_{\rm J}}=4.035\times
E(B-V)$ where we use the dust reddening $E(B-V)$ as a function of
position calculated by Schlegel, Finkbeiner \& Davis
(1998)\nocite{schlegel98}.  We then weight the random distribution
according to the reduction in number density caused by the extinction
such that
\begin{equation}
W_{\rm ext}(\alpha,\delta)=10^{-\beta A_{\rm b_{\rm
J}}(\alpha,\delta)},
\end{equation}
where $\beta$ is the slope of the QSO number counts at the magnitude
limit of the survey.  At $\bj=20.85$, the magnitude limit of the 2QZ,
the QSO number counts are flat, with $\beta\simeq0.3$.  Again we find
that applying this correction makes no significant difference to the
measured $\xi\qso$.

It can be useful to present clustering results in a non-parametric form,
specified by the 
clustering amplitude within a given comoving radius, rather than as a
scale length which depends on a power law fit to $\xi\qso$.  This is
generally represented by the integrated correlation function,
$\xibar$, within a given radius in redshift-space, $s\max$, 
\begin{equation}
\xibar(s\max)=\frac{3}{s\max^3}\int_0^{s\max} \xi(x)x^2{\rm d}x.
\label{xibar_eq}
\end{equation}
Authors tend to choose a variety of values for $s\max$,
e.g. $s\max=10\Mpc$ \cite{sb94,cs96} or $s\max=15\Mpc$ \cite{lac98}.
The choice is a compromise, selecting the scale for
which a significant signal is seen.  It is easiest to relate these
measurements to theory for large scales, where linear evolution
occurs.  Below we will quote clustering amplitudes with $s\max=20\Mpc$
as this is a scale at which evolution should be linear to better than
a few per cent.  We note that choosing a large radius also reduces the
effects of small scale peculiar velocities and redshift measurement
errors, which may well be a function of redshift.

\begin{figure*}
\centering
\centerline{\psfig{file=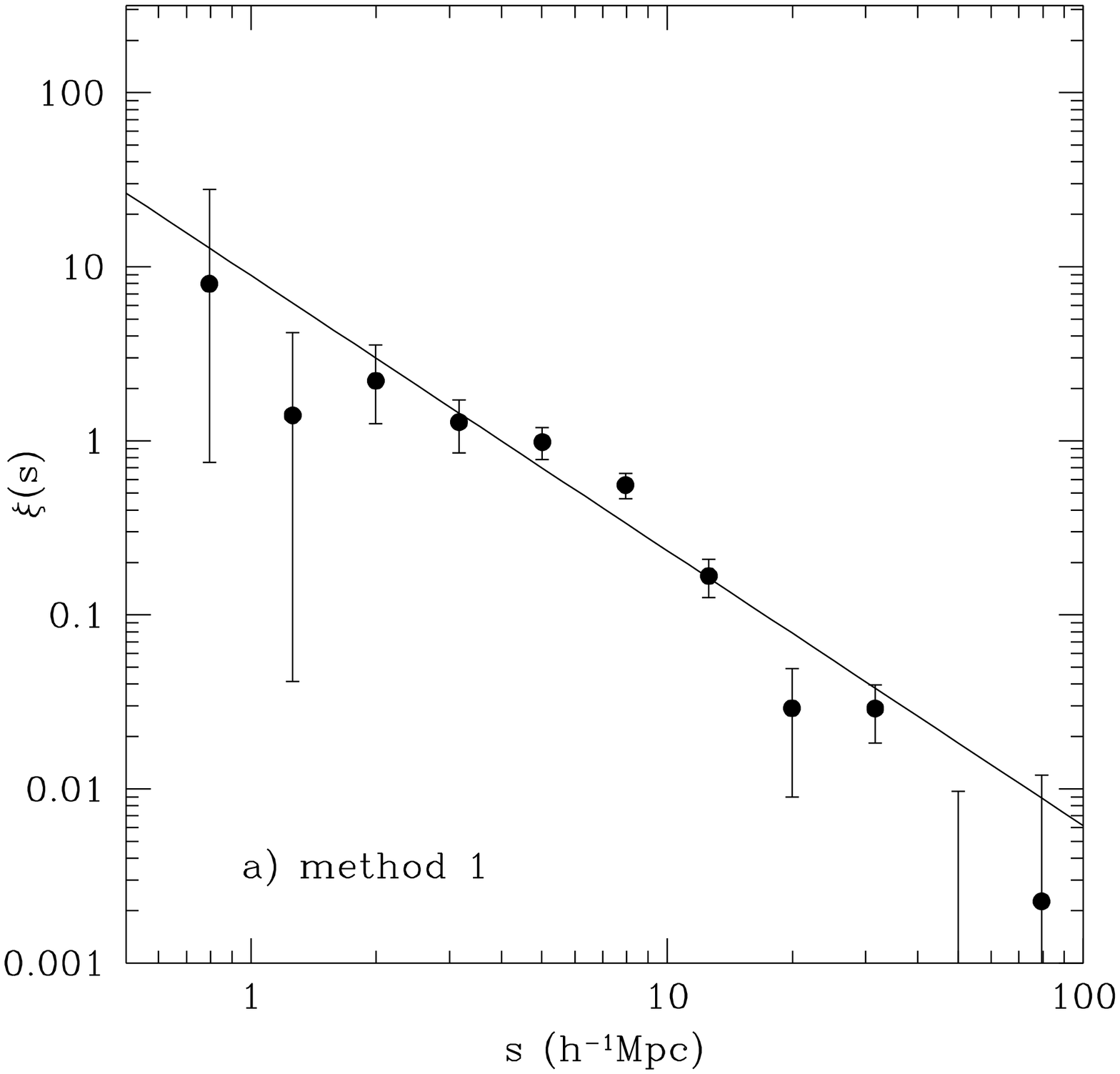,width=8.0cm}\psfig{file=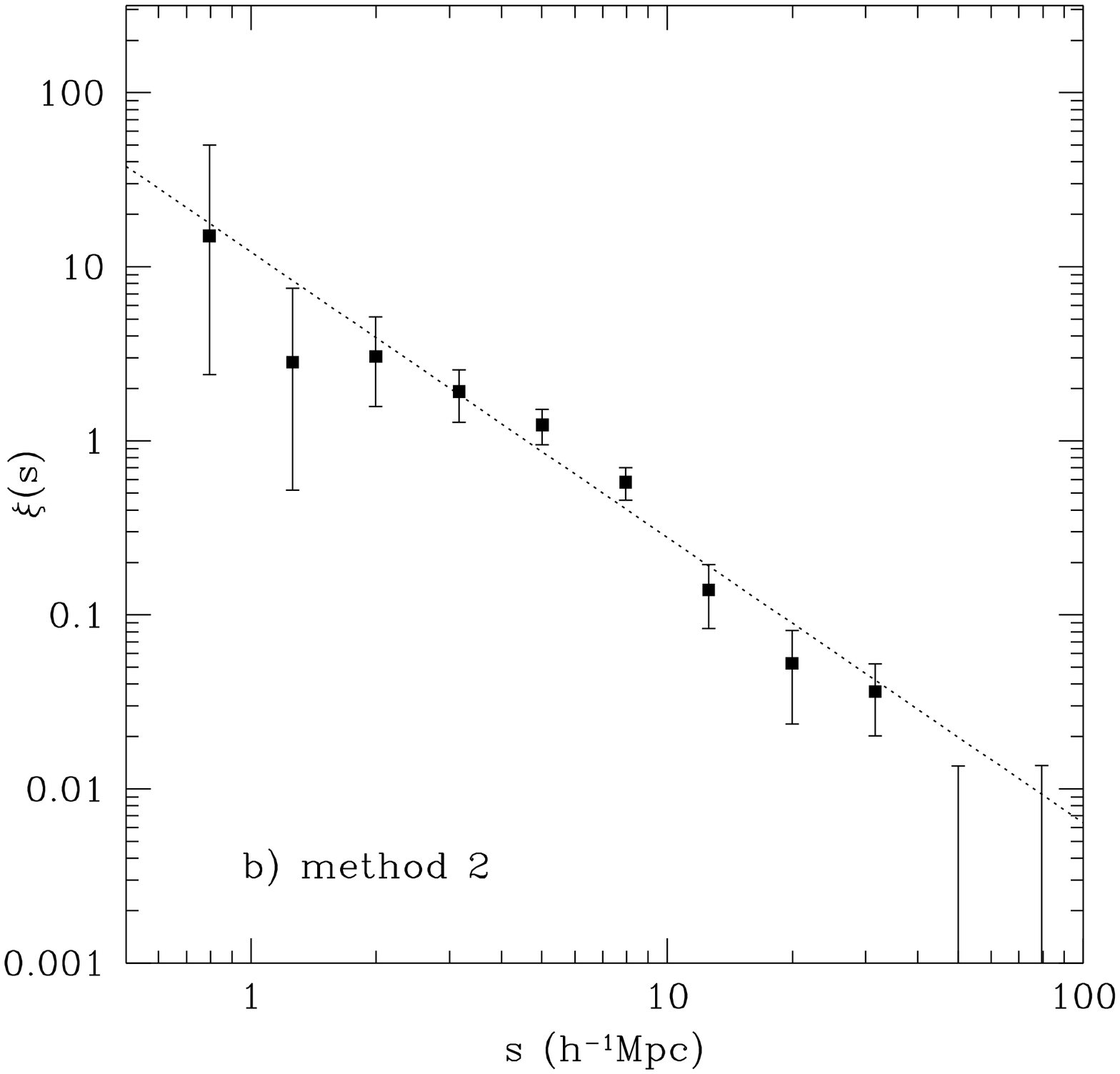,width=8.0cm}}
\centerline{\psfig{file=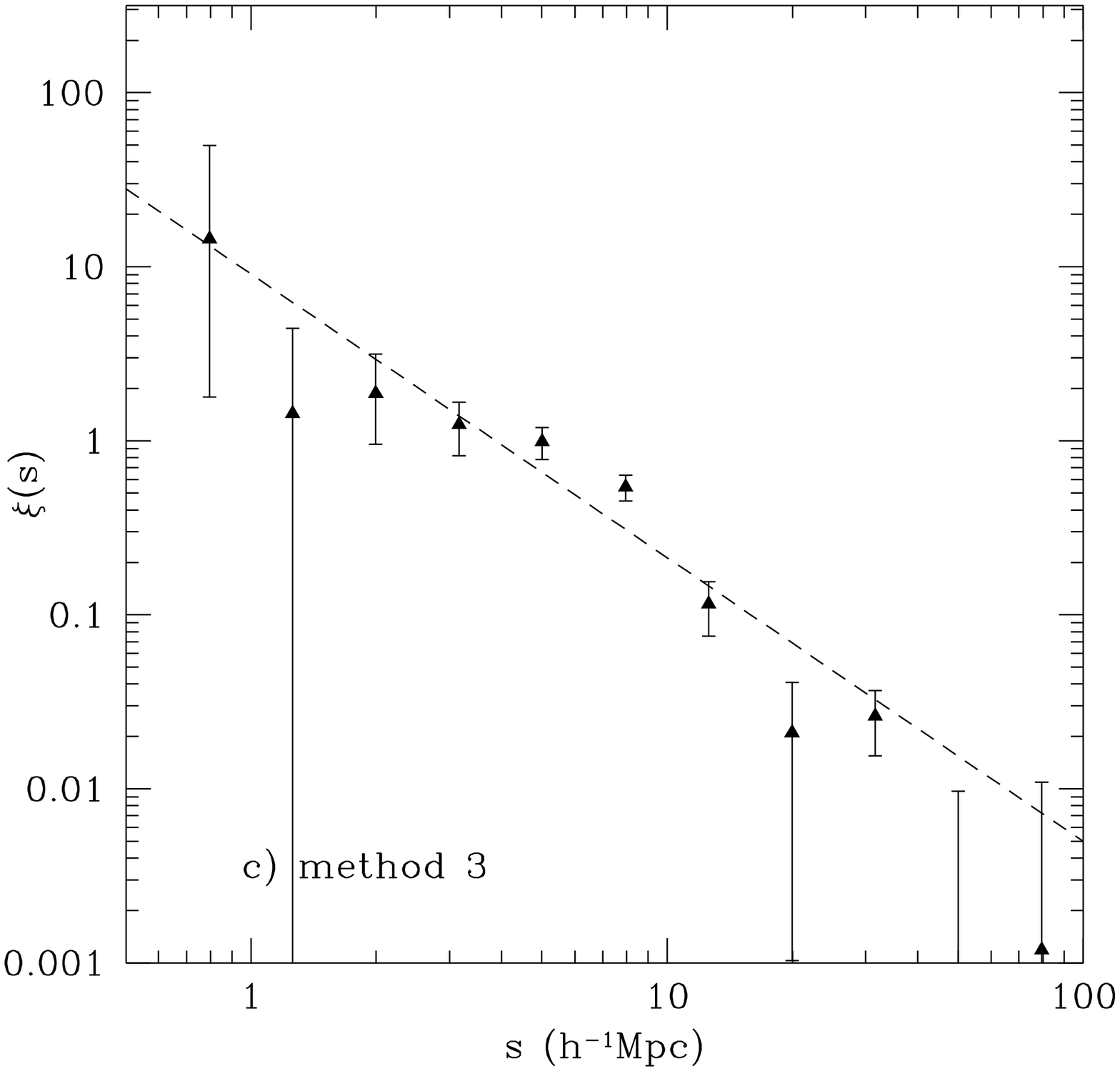,width=8.0cm}\psfig{file=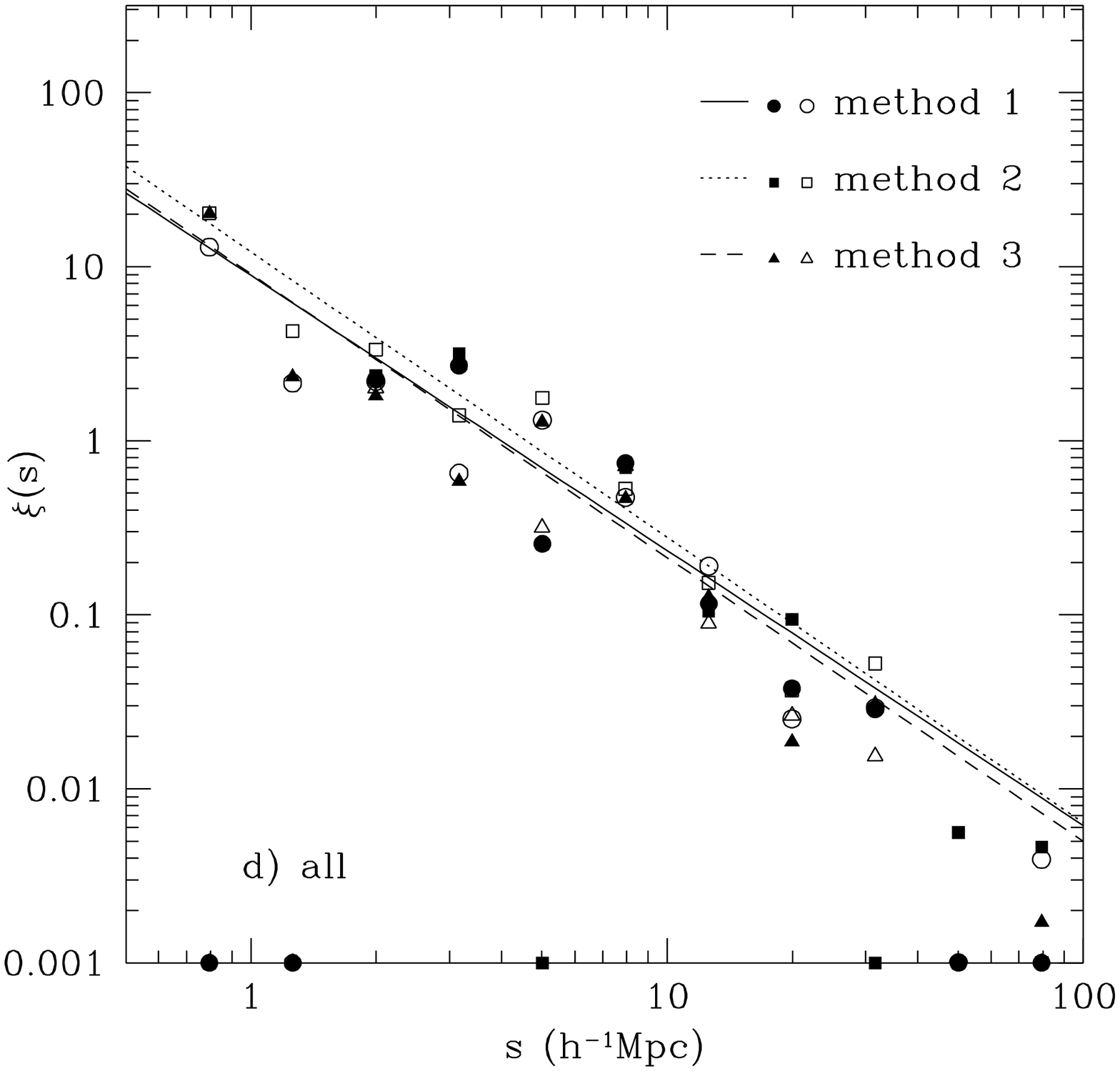,width=8.0cm}}
\caption{The two-point correlation function for 2QZ QSOs in the
redshift interval $0.3<z\leq2.9$ for an EdS cosmology, using the different
estimators discussed in the text. a) method 1, b) method 2, c) method
3 and d) all three methods.  In plots a), b) and c) we show $\xi\qso$
from the combined NGC and SGC strips with the best fit power law at
$s\leq35\Mpc$ in each case.  In d) we compare the three methods, the
two points for each method being separate estimates in the NGC (filled
symbols) and SGC (open symbols) strips.  The lines are identical to
those in a), b) and c) denoting the best fit power law in each method.}
\label{estcomp}
\end{figure*}

We calculate the errors on $\xi\qso$ using the Poisson estimate of
\begin{equation}
\Delta\xi(s)=\frac{1+\xi(s)}{\sqrt{QQ(s)}}.
\label{xierr}
\end{equation}
At small scales, $\lsim50\Mpc$, this estimate is accurate because each
QSO pair is independent (i.e. the QSOs are not generally part of
another pair at scales smaller than this).  On larger scales the QSOs
pairs become more correlated and we use the approximation that
$\Delta\xi(s)=[1+\xi(s)]/\sqrt{N\qso}$, where $N\qso$ is the total
number of QSOs used in the analysis (Shanks \& Boyle 1994; CS96).  In
this paper, we will generally be concerned with analysis on small
scales ($\leq50\Mpc$), where the Poisson error estimates are
applicable.  As a confirmation of our Poisson error estimates we have
also derived field-to-field errors,  by splitting the NGC and SGC
strips into two, and determining the scatter between the resulting
four independent regions.  The errors determined in this fashion are
approximately equal to or less than the Poisson errors.  We also test
bootstrap errors which are found to be $\sim\sqrt{3}$ times greater
than Poisson on all scales of interest, in agreement with expected
theory (Mo, Jing \& Borner 1992) and previous measurements (e.g. Boyle
\& Shanks 1994; CS96).  On small scales, $\lsim2\Mpc$, the number of
QSO-QSO pairs can be $\lsim10$.  In this case simple {\it root-n}
errors (Eq. \ref{xierr}) do not give the correct upper and lower
confidence limits for a Poisson distribution.  We use the formulae of
Gehrels (1986) to estimate the Poisson confidence intervals for
one-sided 84\% upper and lower bounds (corresponding to $1\sigma$ for
Gaussian statistics).  These errors are applied to our data for
$QQ(s)<20$.  By this point root-n errors adequately describe the
Poisson distribution.

\subsection{Fitting models to $\xi(s)$}

Below we make comparisons of the data to a number of models, both
simple  functional forms (power laws) and more complex, physically
motivated, models (e.g. CDM).  We use the maximum likelihood method to
determine the best fit parameters.  The likelihood estimator is based
on the Poisson probability distribution function, so that
\begin{equation}
L=\prod_{i=1}^{N}\frac{e^{-\mu}\mu^{\nu}}{\nu!}
\end{equation}
is the likelihood, where $\nu$ is the observed number of QSO-QSO
pairs, $\mu$ is the expectation value for a given model and $N$ is the
number of bins fitted.  We fit the data with bins $\Delta\log(r)=0.1$,
although we note that varying the bin size by a factor of two makes no
noticeable difference to the resultant fit.  In practice we minimize
the function $S=-2{\rm ln}(L)$, and determine the errors from the
distribution of $\Delta S$, where $\Delta S$ is assumed to be
distributed as $\chi^2$.  This procedure does not give us an absolute
measurement of the goodness-of-fit for a particular model.  We
therefore also derive a value of $\chi^2$ for each model fit in order
to confirm that it is a reasonable description of the data.  In
particular this is appropriate when fitting on moderate to large
scales ($\gsim5\Mpc$), where the pair counts are large enough that the
Poisson errors are well described by Gaussian statistics.

\section{The correlation function of 2QZ QSOs}\label{section_xir}

Here we present the results of our clustering analysis on an initial
sample of 2QZ QSOs.  This sample contains 10558 QSOs taken from
the 2QZ 10k catalogue.  Fig. \ref{wedge} shows the distribution
of QSOs projected onto a plane of constant declination.  We note
that the current distribution is highly non-uniform, as the survey is
only partially complete.

\subsection{The redshift averaged QSO correlation function}

We first measure the QSO two-point correlation function averaged over
the entire redshift interval $0.3<z\leq2.9$.  For an EdS cosmology we
estimate $\xi\qso$ using the three different processes discussed in
Section \ref{xiest}: 1) full accounting for non-uniform coverage, 2)
taking only completely observed regions, and 3) using the QSO
($\alpha$,$\delta$) for the random point positions.  These are
presented in Fig. \ref{estcomp}a, b and c respectively.  The
results demonstrate that QSO clustering follows a power law on small
to intermediate scales.  There is some evidence of a break in the
power law at $\sim35\Mpc$.  We fit a power law of the conventional form, 
\begin{equation}
\xi(s)=\left(\frac{s}{s_0}\right)^{-\gamma}.
\end{equation}
The best fits using the maximum likelihood technique are
$(s_0,\gamma)=(3.99^{+0.28}_{-0.34},1.58^{+0.10}_{-0.09})$,
$(4.59^{+0.37}_{-0.39},1.64^{+0.12}_{-0.12})$ and
$(3.87^{+0.29}_{-0.32},1.63^{+0.11}_{-0.11})$ for methods 1, 2 and 3
respectively, where $s_0$ is in units of $\Mpc$.  We fit the power law
on scales $0.7-35\Mpc$.  The minimum scale is set by the smallest
scale at which we find QSO pairs and the maximum scale is set by
scale of the observed break in $\xi(s)$.
A comparison of all three methods is shown in Fig. \ref{estcomp}d.
Here we also plot separately the clustering of the NGC and SGC
strips. 

\begin{figure}
\centering
\centerline{\psfig{file=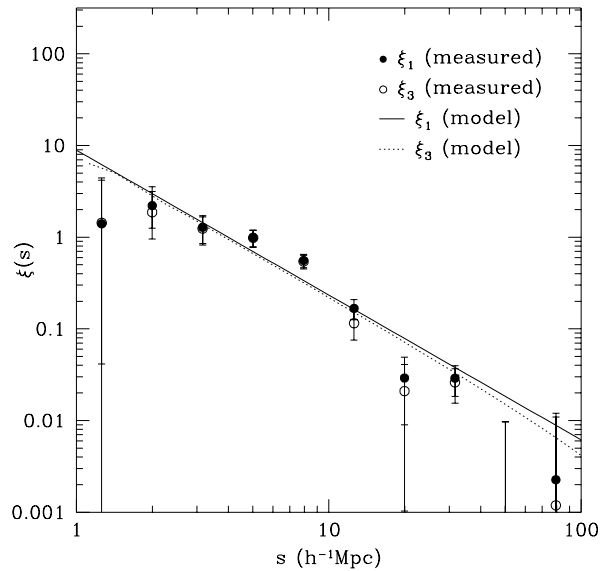,width=8.0cm}}
\caption{The estimated effect of using actual QSO ($\alpha$,$\delta$) for the
random distribution when estimating the correlation function (method
3; dotted line).  The solid line is the input model power law
(identical to the power law fit to method 1).  The filled and open
circles are the estimates of $\xi\qso$ from the data using methods 1
and 3 respectively.}
\label{angsup}
\end{figure}

First, we note that the signals from the NGC and SGC strips are
consistent.  The NGC has no pairs at very small scales ($<1.5\Mpc$),
however the SGC strip only contains 3 pairs at these scales, and fewer
are expected in the NGC due to the smaller number of objects in this
strip (4005 in the NGC vs. 6553 in the SGC).  Second, there appears to
be no significant difference between our different estimations of
$\xi\qso$.  Method 2 shows a slightly higher signal while method 3 is
marginally lower than the other two methods.  We estimate how much of
the difference between methods 1 and 3 could be due to the removal of
real signal by taking the measured correlation function from method 1
and integrating it over our redshift range, weighted by the QSO
redshift distribution.  This then gives us an angular correlation
function with which we weight the random distribution when deriving
the 3-D correlation so suppressing the angular component.  The results
of this analysis are shown in Fig. \ref{angsup}.  On small scales
there is very little effect on the 3-D clustering, however on scales
$\gsim20\Mpc$ the clustering signal becomes suppressed by larger
amounts (dotted line).  The correlation functions measured from the
data using the two methods are also plotted in  Fig. \ref{angsup}.
The difference in the measured values at $\gsim20\Mpc$ is similar to
that predicted by the model, suggesting that some large-scale power is
removed by method 3.  We therefore choose to use method 1 throughout
the remainder of our analysis (method 2 contains half as many QSOs as
method 1, only 5348 and they are generally distributed in many small
overlap regions; the dark shaded regions in fig. \ref{coverage}).  Any
residual systematic errors caused by the variable observational
completeness are not significant enough to affect any of the
conclusions of this paper. 

\begin{figure}
\centering
\centerline{\psfig{file=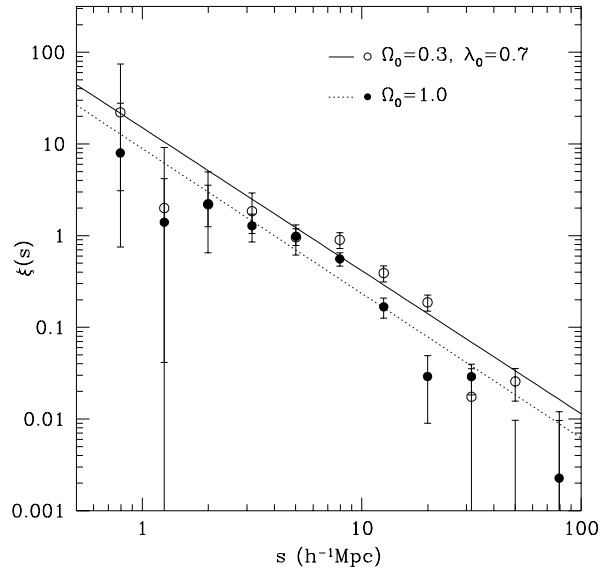,width=8.0cm}}
\caption{A comparison between the 2QZ $\xi\qso$ for different
cosmologies: EdS (filled circles) and $\Lambda$ (open circles).  The
solid and dotted lines are the best fit power laws in each case.}
\label{lamxi}
\end{figure}

\begin{figure*}
\centering
\centerline{\psfig{file=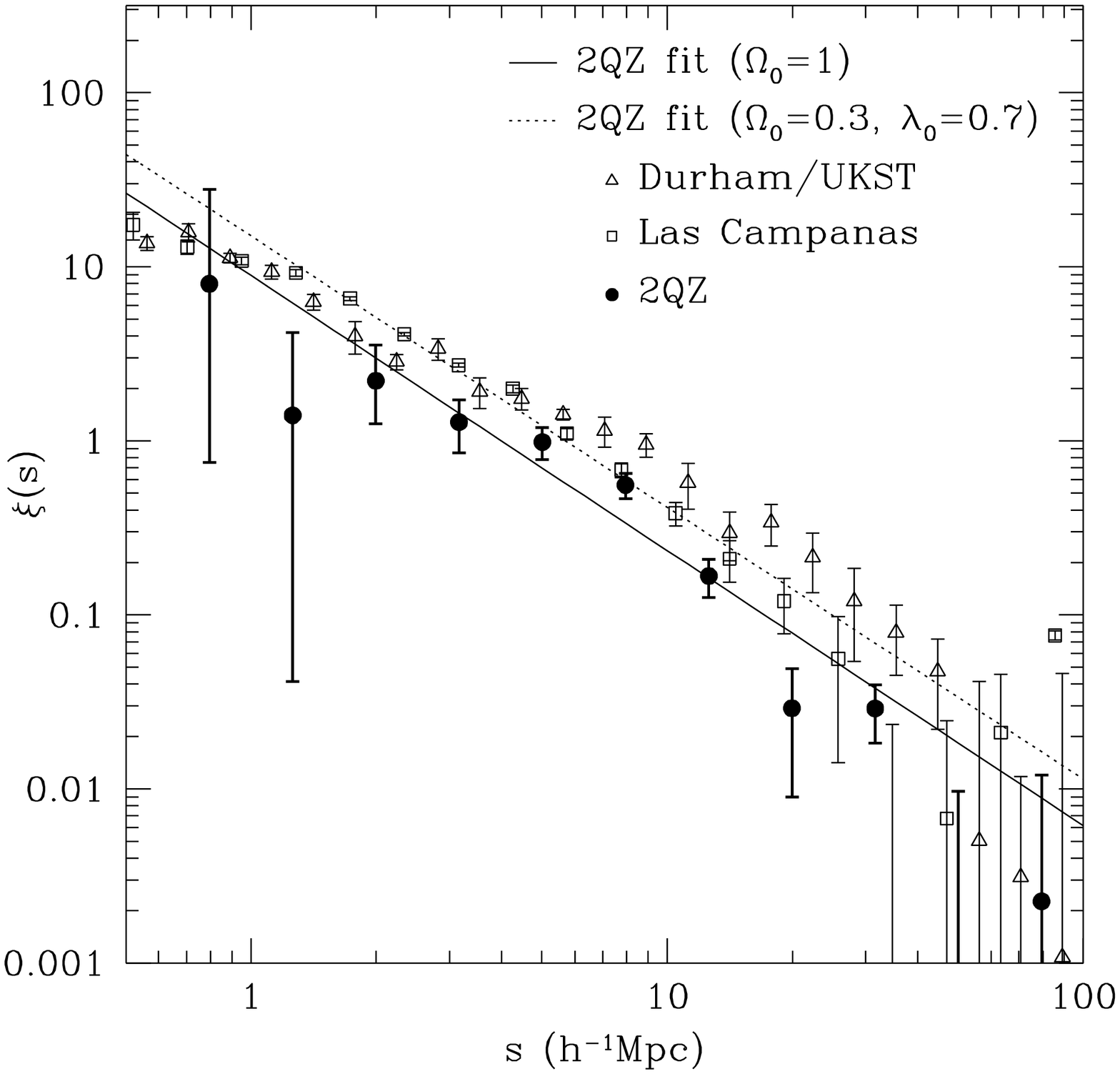,width=8.0cm}\psfig{file=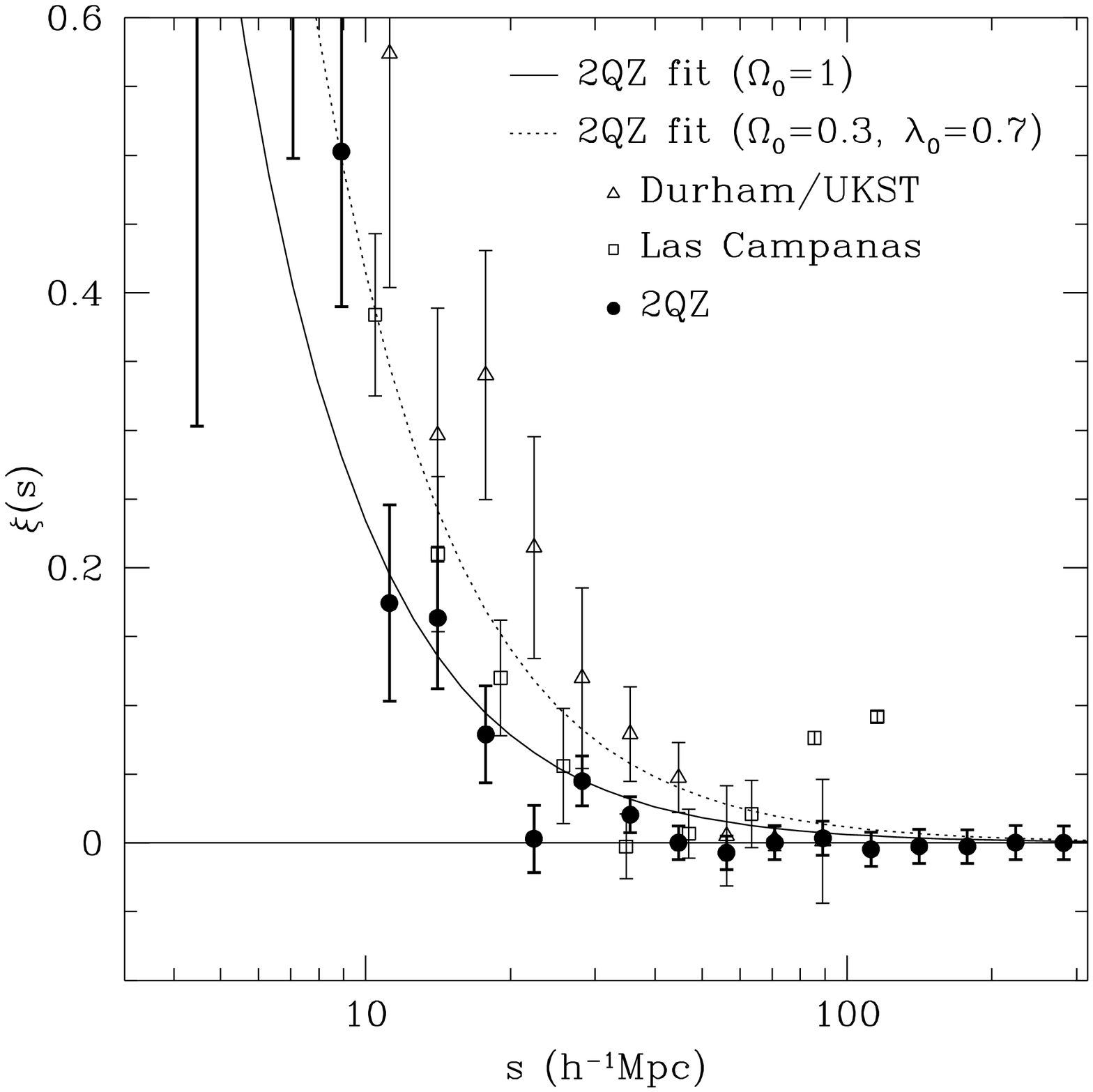,width=8.0cm}}
\caption{The two-point correlation function for 2QZ QSOs in the
redshift interval $0.3<z\leq2.9$ in an EdS cosmology, compared to the
clustering of local galaxies from the Durham/UKST Survey (Ratcliffe et
al. 1998, triangles) and the Las Campanas Survey (Tucker et al. 1997,
squares), plotted in  a) log-log space to highlight smaller scales and
b) log-linear space to highlight larger scales.  In a) the 2QZ data is
plotted in $\log(\Delta s)=0.2$ bins, while in b) $\log(\Delta
s)=0.1$.  The solid line is the best fit power law to the 2QZ data for
the EdS cosmology, while the dotted line is the best fit for the
$\Lambda$ cosmology.  The 2QZ data points for the $\Lambda$
model are omitted for clarity.}
\label{qsogal}
\end{figure*}

\begin{table*}
\baselineskip=20pt
\begin{center}
\caption{2QZ clustering results for various cosmologies and
redshift intervals.  The $s_0$ and $\gamma$ are best fit values.  The
results for the 2 parameter fit are allowing both $s_0$ and $\gamma$
to vary freely.  For the 1 parameter fit we constrain $\gamma$ to be
the best fit value for each cosmology over the full redshift interval
($0.30<z\leq2.90$) and allow only $s_0$ to vary.  The reduced $\chi^2$
for each fit are also listed.}  
\begin{tabular}{cccccccccc}
\hline
&&&& \multicolumn{3}{c}{2 parameter fit} & \multicolumn{2}{c}{1
parameter fit} &\\
($\Om$,$\lo$) & redshift range & $\bar{z}$ & $N\qso$ & $s_0$ &
$\gamma$ & $\chi^2$ &  $s_0$ & $\chi^2$ & $\xibar(20)$\\
\hline
(1.0,0.0) & $0.30<z\leq2.90$ & 1.49 & 10558 & $3.99^{+0.28}_{-0.34}$ & $1.58^{+0.10}_{-0.09}$ & 1.42 & - & - & $0.197\pm0.026$\\
(0.3,0.7) & $0.30<z\leq2.90$ & 1.49 & 10558 & $5.69^{+0.42}_{-0.50}$ & $1.56^{+0.10}_{-0.09}$ & 1.33 & - & - & $0.416\pm0.048$\\
\hline
(1.0,0.0) & $0.30<z\leq0.95$ & 0.69 & 2299 & $3.84^{+0.56}_{-0.69}$ & $1.70^{+0.27}_{-0.36}$ & 1.14 & $3.65^{+0.56}_{-0.56}$ & 1.18 & $0.163\pm0.054$\\
(1.0,0.0) & $0.95<z\leq1.35$ & 1.16 & 2116 & $2.72^{+0.94}_{-1.18}$ & $1.25^{+0.27}_{-0.25}$ & 1.37 & $3.55^{+0.61}_{-0.64}$ & 1.27 & $0.211\pm0.057$\\
(1.0,0.0) & $1.35<z\leq1.70$ & 1.53 & 2177 & $3.49^{+0.61}_{-0.70}$ & $1.63^{+0.22}_{-0.22}$ & 1.51 & $3.41^{+0.56}_{-0.56}$ & 1.53 & $0.192\pm0.052$\\
(1.0,0.0) & $1.70<z\leq2.10$ & 1.89 & 2186 & $4.31^{+0.55}_{-0.61}$ & $1.83^{+0.21}_{-0.20}$ & 0.64 & $3.85^{+0.56}_{-0.56}$ & 0.87 & $0.140\pm0.055$\\
(1.0,0.0) & $2.10<z\leq2.90$ & 2.36 & 1780 & $4.43^{+0.77}_{-0.94}$ & $1.84^{+0.30}_{-0.30}$ & 1.14 & $3.96^{+0.80}_{-0.83}$ & 1.35 & $0.099\pm0.078$\\
\hline
(0.3,0.7) & $0.30<z\leq0.95$ & 0.69 & 2299 & $5.28^{+0.72}_{-0.89}$ & $1.72^{+0.23}_{-0.22}$ & 1.21 & $4.90^{+0.71}_{-0.72}$ & 1.28 & $0.269\pm0.085$\\
(0.3,0.7) & $0.95<z\leq1.35$ & 1.16 & 2116 & $4.05^{+1.21}_{-1.52}$ & $1.38^{+0.27}_{-0.24}$ & 0.69 & $4.65^{+0.89}_{-0.91}$ & 0.63 & $0.371\pm0.102$\\
(0.3,0.7) & $1.35<z\leq1.70$ & 1.53 & 2177 & $5.23^{+0.92}_{-1.08}$ & $1.55^{+0.21}_{-0.20}$ & 1.92 & $5.24^{+0.82}_{-0.81}$ & 1.92 & $0.468\pm0.103$\\
(0.3,0.7) & $1.70<z\leq2.10$ & 1.89 & 2186 & $6.24^{+0.86}_{-0.99}$ & $1.80^{+0.21}_{-0.19}$ & 0.83 & $5.54^{+0.84}_{-0.86}$ & 1.09 & $0.394\pm0.110$\\
(0.3,0.7) & $2.10<z\leq2.90$ & 2.36 & 1780 & $6.93^{+1.32}_{-1.64}$ & $1.64^{+0.29}_{-0.27}$ & 1.26 & $6.68^{+1.23}_{-1.27}$ & 1.30 & $0.615\pm0.178$\\
\hline
\label{clusres}
\end{tabular}
\end{center}
\end{table*}

\begin{figure}
\centering
\centerline{\psfig{file=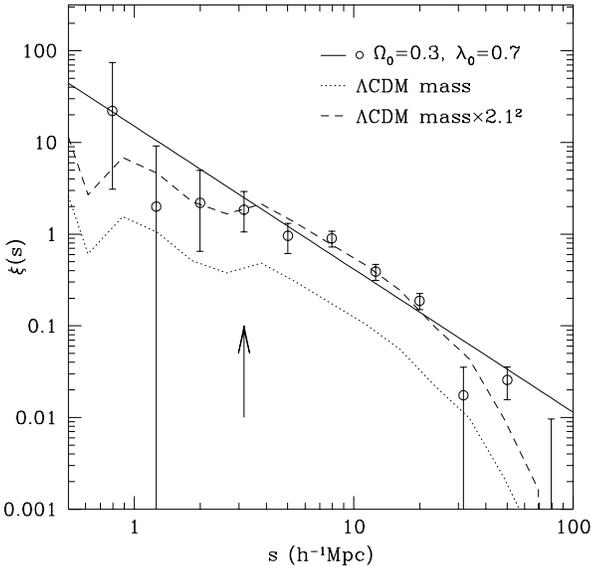,width=8.0cm}}
\caption{A comparison between the 2QZ $\xi\qso$ (open circles) and
$\xi_\rho$ (dotted line) from the Hubble Volume simulations in the $\Lambda$
cosmology.  The dashed line is $2.1^2\times\xi_\rho$.  The
arrow marks the resolution limit of the simulation.  Also shown
is the best fit power law (solid line).}
\label{hubvolxi}
\end{figure}

\begin{figure*}
\centering
\centerline{\psfig{file=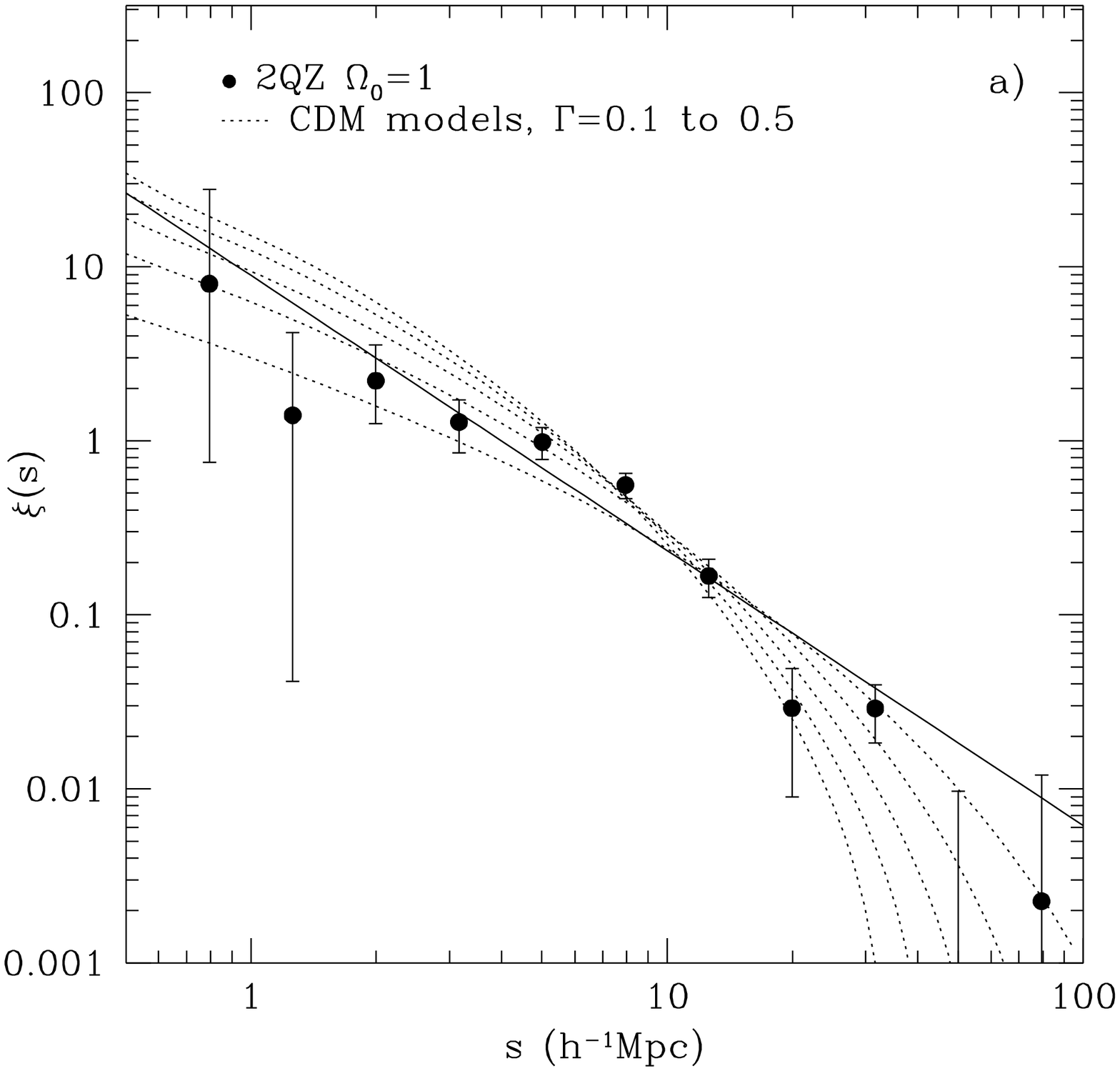,width=8.0cm}\psfig{file=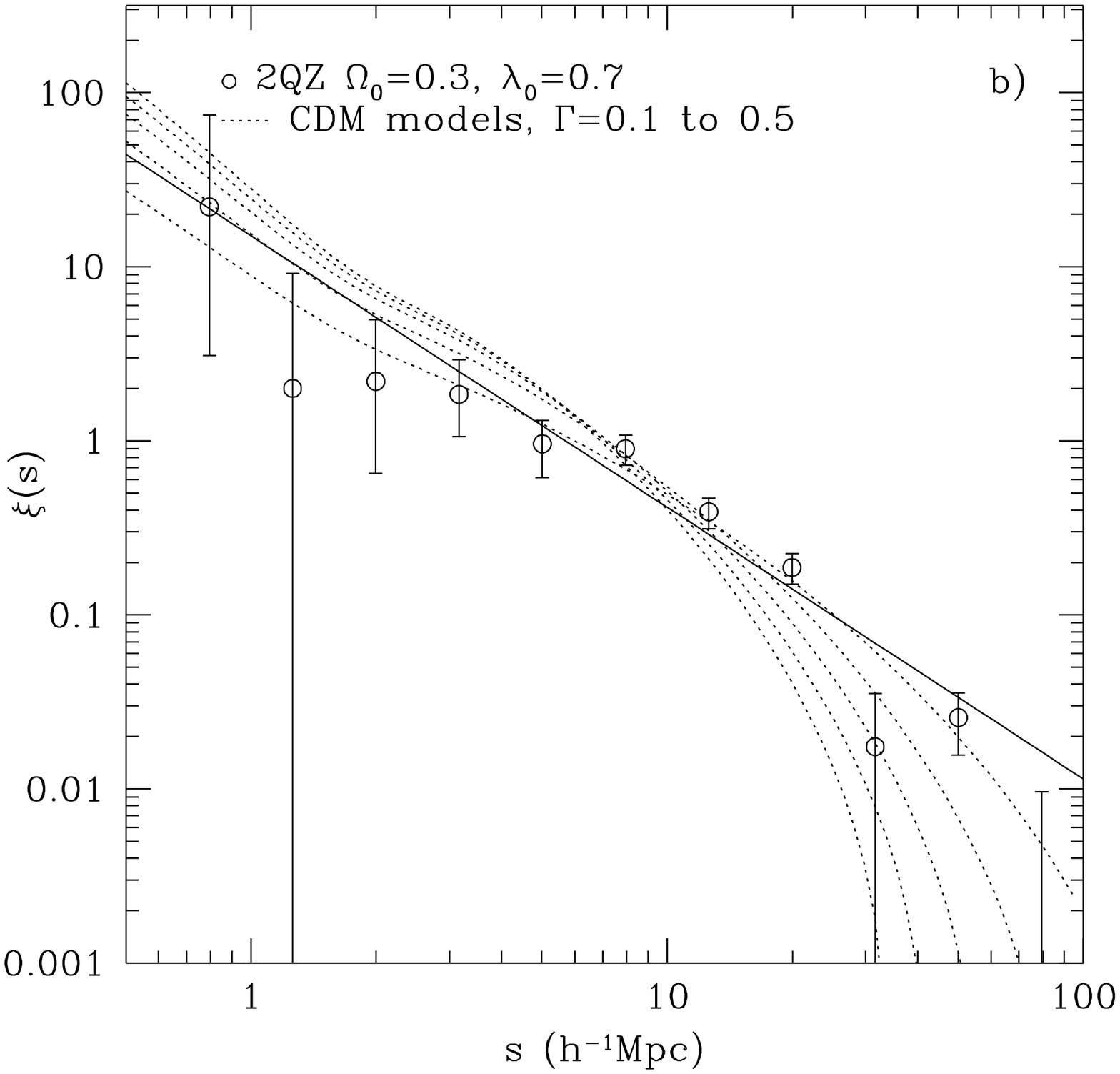,width=8.0cm}}
\caption{A comparison between the 2QZ $\xi\qso$ and and CDM type
models in the a) EdS and b) $\Lambda$ cosmologies.  The dotted lines
are non-linear CDM correlation functions with various values of
$\Gamma_{\rm eff}$ scaled by a linear bias to give the best fit to the
2QZ data. $\Gamma_{\rm eff}=0.1$, 0.2, 0.3, 0.4 and 0.5 from top to
bottom at large scales).}
\label{xircdm}
\end{figure*}

We also calculate $\xi\qso$ for the $\Lambda$ cosmology, with
$\Om=0.3$ and $\lo=0.7$.  This is compared to the method 1
estimate for the EdS case in Fig \ref{lamxi}.  The effect of
introducing a significant cosmological constant term is to increase
the relative separation of QSOs, and hence increase the clustering
scale length.  The break in the power law is now seen at
$\simeq60\Mpc$, we therefore make our power law fits out to this
scale.  The best fit power law for the $\Lambda$ cosmology is
$(s_0,\gamma)=(5.69^{+0.42}_{-0.50},1.56^{+0.10}_{-0.09})$.  All
results are listed in Table \ref{clusres}. 

\subsection{QSO clustering compared to local galaxies}

In Fig. \ref{qsogal} we compare our QSO clustering results at
$\bar{z}=1.49$ to galaxy clustering at low redshift ($z\sim0.05$).  In
particular the Las Campanas \cite{tucker97} and Durham/UKST
\cite{rspf98} galaxy surveys (open squares and triangles
respectively).  We see that there is good general agreement between
the galaxy and QSO clustering, although the samples have differing
redshift ranges.  The EdS $\xi\qso$ is slightly lower on average than
$\xi\gal$, while $\xi\qso$ in the $\Lambda$ cosmology is closer in
amplitude to the galaxies.  Both QSOs and galaxies show a break in
$\xi(r)$ at $\sim40\Mpc$.  We note that the errors on $\xi\qso$ are
smaller than those on $\xi\gal$ at $>20\Mpc$.

\subsection{QSO clustering compared to CDM}

In order to compare directly to theory, and include all non-linear
effects and redshift-space distortions, we have used the Hubble Volume
simulations of the Virgo Consortium \cite{hubvol98}.  We have produced
mock 2QZ QSO catalogues with the same survey geometry and explicitly
included evolution of the density field by outputting the simulation
at different times along the light cone.  A detailed discussion of the
simulations, including a number of biasing models will be given in
Hoyle et al. (2001 in preparation).  Here we simply compare the dark matter
correlation function averaged over the light cone to the 2QZ data.  In
particular we compare the redshift-space mass correlation function of
a $\Lambda$CDM model to  $\xi\qso$.  This model has $\Om=0.3$,
$\Omega_{\rm baryon}=0.04$, $\lo=0.7$, $\sigma_8=0.9$, $h=0.7$ and an
effective shape parameter, taking into account the baryon component
\cite{sug95}, of $\Gamma_{\rm eff}=0.17$.  The model and data are
shown in Fig. \ref{hubvolxi}.  The  amplitude of $\xi\qso$ is $\sim4$
times larger than the $\Lambda$CDM mass correlation function,
$\xi_\rho$.  When scaled by this factor, the model and data appear to
be well matched with a best fit bias value of 2.1.

Increasing the value of $\Gamma_{\rm eff}$ will move the models away
from the data by steepening $\xi_{\rho}$ at large scales.  We do not
have a large suite of simulations with which to compare the effect of
changing $\Gamma_{\rm eff}$ and cosmology.  However, on the scales
which we are fitting, linear theory is a reasonable approximation.
Therefore the effect of redshift space distortions will be simply to
scale $\xi$ by $(1+2\beta/3+\beta^2/5)$ where $\beta\simeq\Om^{0.6}/b$
\cite{kaiser87}.  We can then simply absorb this factor into an
effective linear bias factor.  We then fit model real space non-linear
correlation functions at $z=1.49$ to the data (again at scales 5 to
$100\Mpc$) using the  ansatz of Peacock \& Dodds (1996) to determine
the non-linear correction to the model $\xi$.  The deviation from
non-linearity is small (typically $\lsim5\%$) on the scales of
interest.  We do not take into account small-scale non-linear velocity
dispersions in our model, however these should be small at the scales
and redshifts considered.  We also do not consider the effects of
redshift measurement errors on $\xi(s)$, these again should only be a
factor on small, $\lsim5\Mpc$, scales.  We use five different models
with $\Gamma_{\rm eff}=0.1$, 0.2, 0.3, 0.4, 0.5 and fit for the
effective linear bias value.  The results of this procedure are shown
in Fig. \ref{xircdm}.  In the EdS cosmology, models with $\Gamma_{\rm
eff}=0.2$, 0.3 and 0.4 are  acceptable at the 10\% level, while
$\Gamma_{\rm eff}=0.1$ and 0.5 are ruled out at greater than $90\%$
confidence.  The main reason that a broad range of models are
acceptable is the relatively low point at $\sim20\Mpc$ in $\xi\qso$.
In  the $\Lambda$ cosmology the $\Gamma_{\rm eff}=0.1$ and 0.2 models
are the only ones to agree with the data, the others being ruled out
at greater than 99.9\% confidence.  Thus the QSO correlation function
detects excess large-scale power over what is expected in the
$\Gamma_{\rm eff}=0.5$ {\it standard} CDM model.  Confirming the
results from the APM galaxy survey \cite{mesl90}.  

The required
$\Gamma_{\rm eff}$ is larger in  the $\Lambda$ cosmology, as structure
is moved to larger scales.  This suggests a test with the full 2QZ
which will be devoid of observational incompleteness as well as having
increased statistical accuracy.  The break of the correlation function
in a CDM type cosmology can be used as a standard rod to determine
cosmological parameters, in particular $\lo$, if it is at linear
scales.  For example, if at low redshift the shape is well defined,
then if the break is in the linear theory regime it should remain at
the same scale at high redshift.  Measuring the break at a different
scale at high redshift would imply the wrong cosmological parameters
were being used in the determination of the high redshift correlation
function.  This is similar to the geometric tests discussed by several
authors \cite{ap79,p94,bph96} but has the advantage of not being
affected by redshift-space distortions if clustering can be measured
on a sufficiently large scale.  This is because linear redshift space
distortions only affect the amplitude and not the shape of $\xi$.
Shanks \& Boyle (1994) proposed a similar method, using linear
features in the correlation function on $\gsim100\Mpc$ scales.

\begin{figure*}
\centering
\centerline{\psfig{file=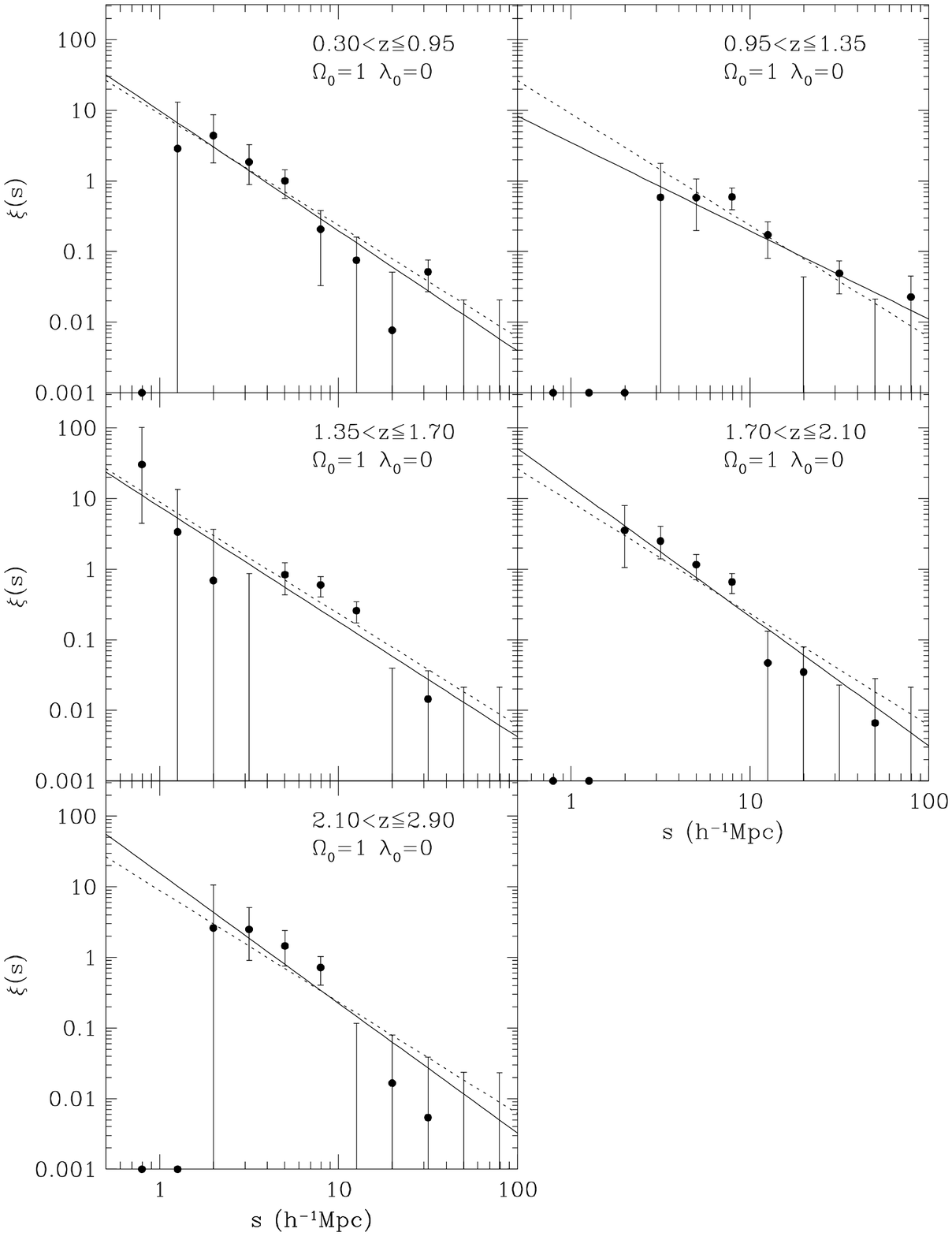,width=16.0cm}}
\caption{The two-point correlation function for 2QZ QSOs as a function
of redshift for the EdS cosmology.  Redshift increases, left to
right and top to bottom.  In each plot the solid line is the best fit
power law on scales $\leq35\Mpc$.  The dotted line is the best fit to
all the QSOs in the redshift range $0.3<z\leq2.9$ and is shown to
aid comparison between redshift intervals.  The points without error
bars at $\xi(s)=0.001$ are where there are zero QSO pair counts in a bin.
These points are fully taken into account in the fitting process.}
\label{xirevol}
\end{figure*}

\begin{figure*}
\centering
\centerline{\psfig{file=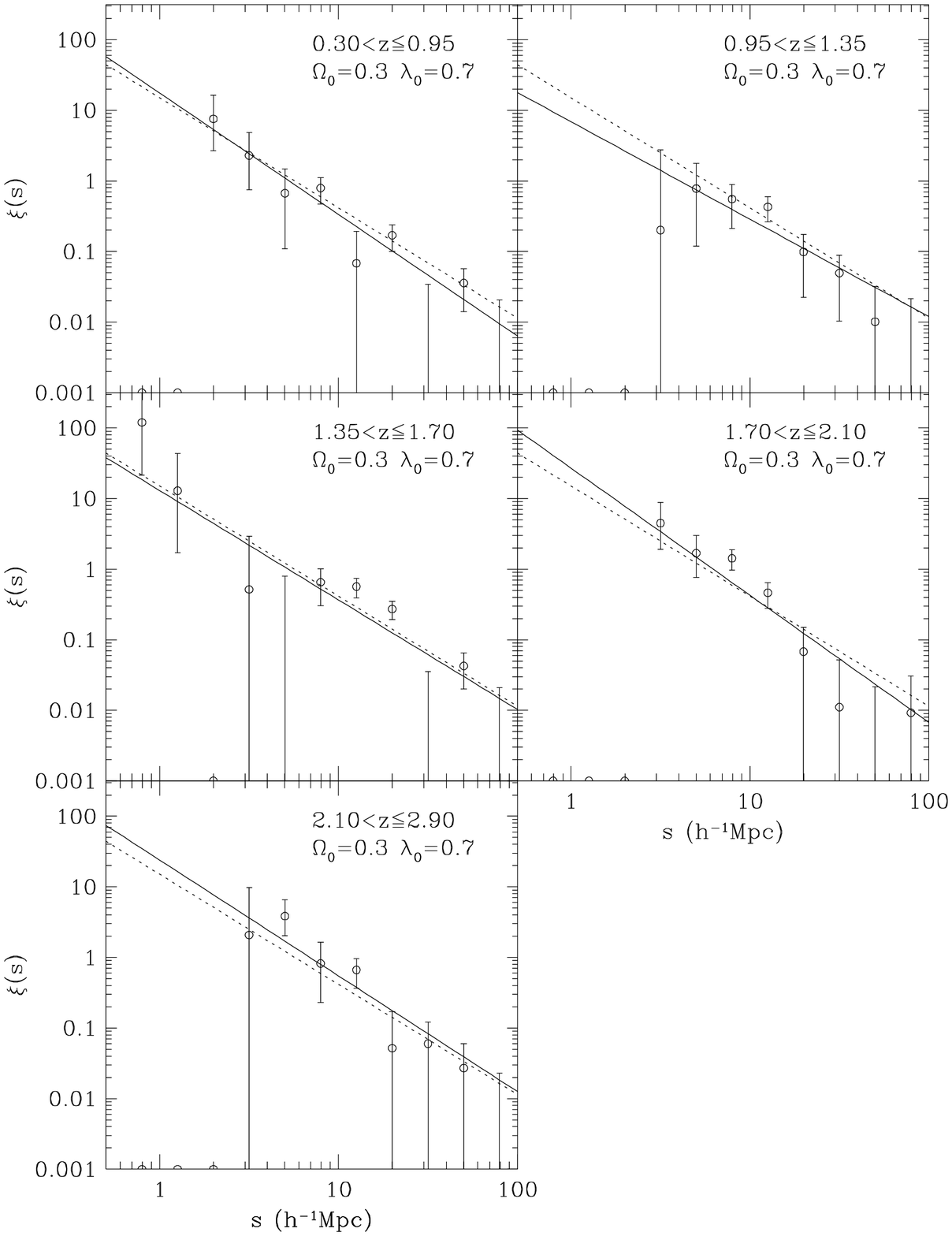,width=16.0cm}}
\caption{The two-point correlation function for 2QZ QSOs as a function
of redshift for the $\Lambda$ cosmology.  Redshift increases, left to
right and top to bottom.  In each plot the solid line is the best fit
power law on scales $\leq60\Mpc$.  The dotted line is the best fit to
all the QSOs in the redshift range $0.3<z\leq2.9$ and is shown to
aid comparison between redshift intervals.}
\label{xirevollam}
\end{figure*}

\begin{figure*}
\centering
\centerline{\psfig{file=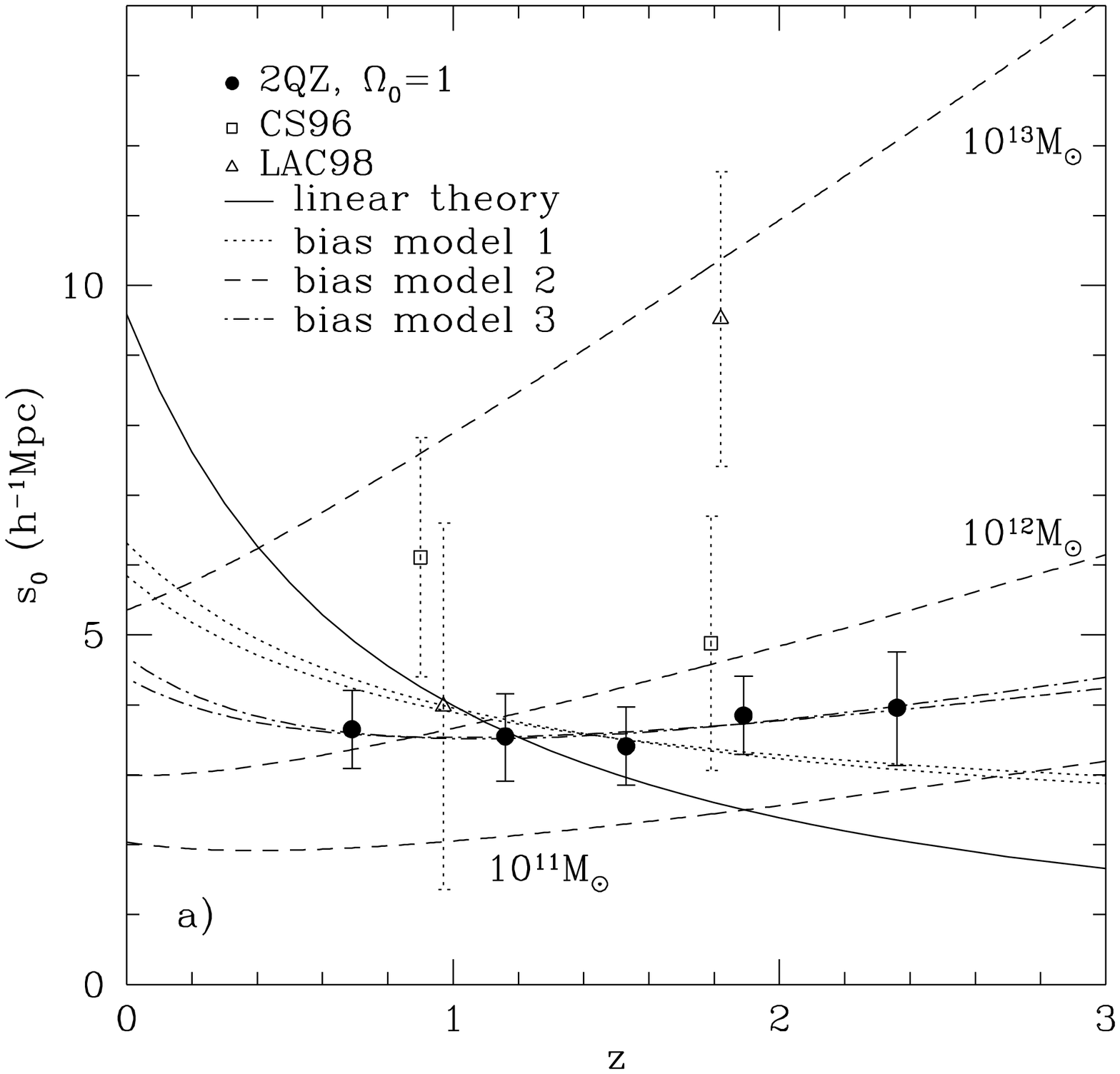,width=8.0cm}\psfig{file=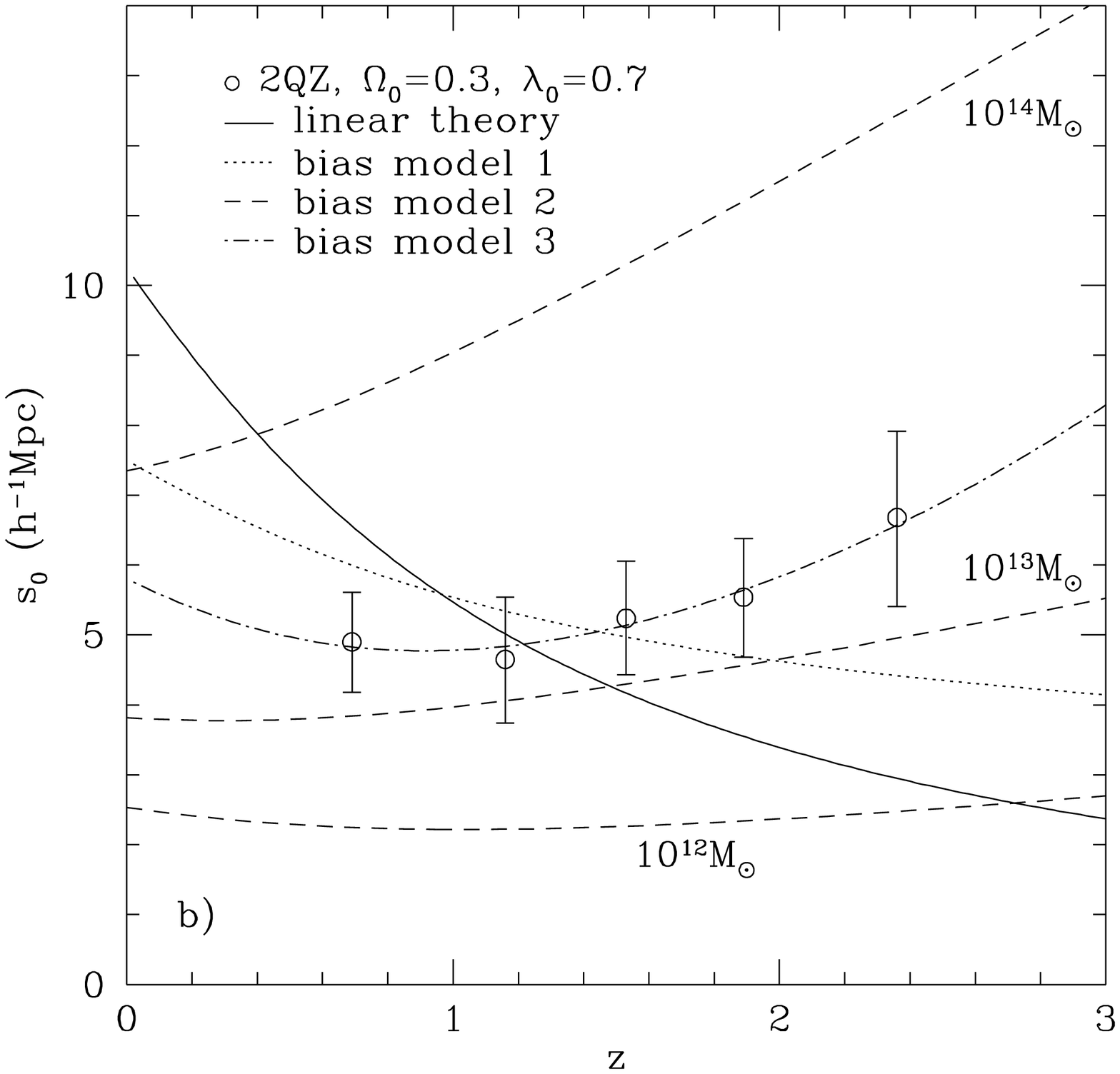,width=8.0cm}}
\caption{The best fit values for $s_0$ with fixed $\gamma$ as a
function of redshift in the a) EdS and b) $\Lambda$ 
cosmologies.  The solid lines show the best fit linear theory
evolution model in each case.  The dotted lines show the best fit
long-lived QSO biasing model (bias model 1) from Section
\ref{longlived}.  The dashed lines show the the models of Matarrese et
al. (1997) (bias model 2) for different values of the minimum halo
mass.  The dot-dashed lines show the best fit empirical bias model
(bias model 3).  In a) the two dotted and dot-dashed lines are for
{\it COBE} (top at $z=0$) and cluster (bottom at $z=0$) normalization.}
\label{r0evol}
\end{figure*}

\section{Evolution of QSO clustering}\label{section_evol}

\subsection{Measurements of QSO clustering evolution}

In the previous section we calculated $\xi\qso$ averaged over a large
redshift interval.  We now split the 2QZ QSO sample up into five
redshift intervals containing approximately equal numbers of QSOs.
The exact limits and numbers of QSOs are given in Table \ref{clusres}.
The measured $\xi\qso$ are shown in Fig. \ref{xirevol} for the EdS
cosmology.  QSO clustering appears to vary little over the entire
redshift range we consider.  The data points are consistent with the
redshift averaged $\xi\qso$ (dotted line in Fig. \ref{xirevol}).  For
each redshift interval we fit a power law, the results of which are
shown by the solid lines in Fig. \ref{xirevol} and in Table
\ref{clusres}.  As for the redshift averaged analysis we fit the power
law within $35\Mpc$.  We similarly fit $\xi\qso$ in redshift intervals
for the $\Lambda$ cosmology  (Fig. \ref{xirevollam}) using the
$60\Mpc$ maximum as above.  Again there is very little evidence of
evolution.  We note that there is some variation in the slope and
amplitude of these power laws, but this appears to be mainly driven by
the relatively low signal-to-noise in each redshift bin.  Great care
should be taken when trying to interpret these power law fit results,
as amplitude and slope are correlated.

An alternative method to derive a measurement of evolution is to
constrain the power law slope and fit only for the scale length,
$s_0$.  This should be valid as we don't see any evidence for
significant evolution in the slope of $\xi\qso$.  We constrain the
slope to be that found over the full redshift range (Section
\ref{section_xir}), $\gamma=1.58$ for the EdS cosmology and
$\gamma=1.56$ for the $\Lambda$ cosmology.  The results of this
fitting process are seen in Table \ref{clusres} and plotted in
Fig. \ref{r0evol}.  In Table \ref{clusres} we also list the reduced
$\chi^2$ values for these fits.  Limiting the fit to one parameter
does not significantly alter the $\chi^2$ values, demonstrating that
the redshift averaged power law slope is a reasonable description of
the data at all redshifts.  Fig. \ref{r0evol}a shows that in the EdS
cosmology clustering is constant as a function of redshift.  The
$\Lambda$ cosmology result is shown in Fig. \ref{r0evol}b.  In this
case there appears to be a marginal increase by a factor of $\sim1.4$
from $z=0.7$ to $2.4$.  We compare our results in the EdS cosmology to
previous QSO clustering results from CS96 and LAC98, using their
measurements of $\xibar$ from 10 and $15\Mpc$ to obtain a value of
$s_0$ assuming a $\gamma=1.58$ power law (the best fit power law
slope).  Our results are in disagreement with those of LAC98 who find
a $\sim2\sigma$ increase in clustering between $z=0.95$ and $z=1.8$.
A possible cause of this is {\it cosmic variance} as LAC98 carry out
their analysis in a single 24.6 deg$^2$ area of sky.  However, given
the large errors on the LAC98 data points, they only disagree with the
2QZ results at $\sim2\sigma$ at $z=1.8$.

A non-evolving clustering distribution has strong implications for
models of structure and QSO formation.  We first compare the 2QZ data
to the simplest possible model, that of linear theory gravitational
evolution in an $\Om=1$ universe.  This model is applicable when QSOs
either directly trace the mass distribution, or have a bias which is
constant as a function of redshift.  When fitting linear theory to the
evolution in $s_0$ for the EdS cosmology we find that the model is
rejected by the 2QZ data at 99.8 per cent confidence.  In the
$\Lambda$ cosmology the linear theory evolution rate is reduced.
However in Fig. \ref{r0evol}b we see that $s_0$ increases with
redshift, although the significance of the increase is marginal: a
constant $s_0$ as a function of redshift is not rejected by the data.
When we try to  fit linear evolution in this case it is  rejected at
$>99.9$ per cent significance.  If we require that the normalization
of the mass clustering be fixed by either the local abundance of
massive clusters \cite{ecf96} or the 4-yr \cobe results \cite{cobe96}
then the mass clustering scale length is forced to be less than
$s_0(z=0)\sim5\Mpc$.  In this case linear theory evolution is even
more clearly rejected by the 2QZ data.  It therefore appears that QSO
clustering cannot follow the linear evolution of the density field,
and QSO bias must be a function of redshift.

We should also make comparisons to galaxy clustering measurements.
The typical scale length found in local galaxy surveys is
$s_0\sim5-6\Mpc$, only marginally higher than the 2QZ results for the
EdS cosmology, and identical to the values found in the $\Lambda$
cosmology.  At $z\sim3$ Adelberger et al. (1998) find a scale length
of $r_0\sim4-6\Mpc$ for Lyman-break galaxies, depending on the assumed
cosmology.  This again is very similar to the results derived from the
2QZ.

\subsection{Comparison to biased models of clustering evolution}

In the previous section we showed that for viable cosmological models,
with evolution based on the gravitational growth of structure, QSOs do
not simply trace the density fluctuations in the Universe.  Therefore
QSOs are related to the mass distribution via a redshift dependent
bias.  The form of this bias depends on the physical mechanisms of QSO
formation.

The question of QSO lifetimes can be linked to their clustering.  If
QSOs have lifetimes which are cosmologically long ($\sim$ a Hubble
time), this would imply that QSOs are intrinsically rare.  They could
therefore be highly clustered, existing in  rare high peaks in the
density field \cite{er88}, assuming that halo mass is the dominant
factor in QSO formation.  However, QSOs could form in a `random'
subset of less biased haloes, with the formation being driven by
mechanisms other than mass, e.g. angular momentum.  

Alternatively, QSOs could have shorter lifetimes, of the order
$\sim10^6-10^8$ years.  There is mounting evidence for this, with the
suggestion that most nearby galaxies appear to contain central
supermassive black holes (e.g. Magorrian et al. 1998), so that most
galaxies pass through a QSO/AGN phase.  They could therefore be
clustered in a similar manner to galaxies.  However, even if all
galaxies go through a QSO/AGN phase, it is possible that this phase
picks out a particular time in the evolution of galaxies, e.g. epochs
of major star formation or merging.  QSO clustering evolution can
potentially help us to distinguish between a number of possible QSO
formation mechanisms.  However, we should be wary of over interpreting
models which do not include the uncertain physical mechanisms required
for QSO formation.

\subsubsection{A long-lived QSO model}\label{longlived}

The next simplest assumption, after assuming that bias does not evolve
with redshift, is that QSOs are long lived (with ages of order a Hubble
time).  We assume that after 
formation at some arbitrarily high redshift the subsequent evolution
of QSO clustering is governed purely by their motion within the
gravitational potential produced by the density fluctuations in the
Universe \cite{f96}.  This then implies a bias which evolves as
\begin{equation}
b(z)=1+(b(0)-1)G(\Om,\lo,z).
\label{biasmodel_1}
\end{equation}
We call this {\it bias model 1}.  $G(\Om,\lo,z)$ is the linear growth
rate of density perturbations,  which for an EdS cosmology is $1+z$.
For the cosmological dependence of the growth rate we use the accurate
fitting formula of Carroll, Press \& Turner (1992), which is good to a
few per cent (note that our $G(\Om,\lo,z)$ is the full evolution term,
and shouldn't be confused with the function of Carroll et al. which
only contains the cosmological dependence).  The biasing model of
Eq. \ref{biasmodel_1} is also equivalent to QSOs forming in peaks of
the density field above a constant threshold (CS96).  This model
places certain limitations on the form of evolution.  First, bias will
tend to unity as time increases.  Secondly, positive evolution (an
increase in clustering) as redshift increases is not possible.  This
is because at most the bias only evolves as fast as $G(\Om,\lo,z)$,
cancelling out the growth in the density field. 

For comparison to the observed clustering we have normalized the mass
evolution in two ways; using both local cluster abundances
\cite{ecf96} and the 4-yr \cobe  results \cite{cobe96}.  We calculate
$s_0(z)$ for the mass assuming a CDM power spectrum with a shape
parameter of $\Gamma_{\rm eff}=0.25$ (varying the shape parameter
$\Gamma_{\rm eff}$ only has an impact on the normalization when using
the \cobe data).  In the EdS cosmology the $s_0$ fits give
$b(0)=1.82^{+0.07}_{-0.07}$ ($1.62^{+0.07}_{-0.06}$) for cluster ({\it
COBE}) normalization.  These correspond to $\sigma_8\simeq1$ for QSOs
at $z=0$ ($\sigma_8$ for mass fluctuations is 0.52 and 0.65 for
cluster and \cobe normalization respectively), which is the same as
the nominal $\sigma_8\sim1$ value found for local galaxies.

Schade, Boyle \& Letawsky (2000) find that at low redshift typical
QSOs and AGN (where by typical we mean at or around the break in the
luminosity function) have host galaxies that are remarkably similar to
normal galaxies, except for a bias towards spheroid dominated
galaxies.  Approximately 55 per cent of their sample had hosts which
were best fit by a bulge-only model.  Elliptical galaxies are well
known to be more strongly clustered than spirals \cite{lmep95} with a
relative bias factor of $b_{\rm e,s}\simeq1.9$.  Correcting for this
morphological segregation gives an expected $\sigma_8=1.2-1.3$ for
QSOs at low redshift, approximately in line with the above value.

For the $\Lambda$ cosmology a biased model of the form in
Eq. \ref{biasmodel_1} provides only a marginally adequate fit to the
data (rejected at 88\%) with a best fit bias of
$b(0)=1.84^{+0.08}_{-0.08}$.  This is because the model cannot
reproduce the increase in clustering strength at high redshift visible
in this cosmology.  The hypothesis that QSOs have cosmologically long
($\sim$ Hubble time) lifetimes therefore appears unlikely in the
$\Lambda$ cosmology.

\subsubsection{More general models of biasing}

The above simple model of biasing can be extended in a number of ways.
The most obvious is to remove the constraint that objects formed at an
arbitrarily high redshift, and allow objects to continue to form at
lower redshift.  The problem then becomes one of deciding how and when
objects do form.  A natural method for deciding when dark matter haloes
form is based on an application of the  Press-Schechter (1974)
formalism which describes the evolution of the number density of dark
matter haloes.  Working within this formalism Mo \& White (1996) have
obtained an approximation for the linear bias of dark matter haloes as
a function of mass.  Matarrese et al. (1997) have used these ideas to
provide biasing models in a {\it COBE} normalized $\Om=1$ universe
assuming a CDM power spectrum with a shape parameter of $\Gamma_{\rm
eff}=0.25$. 
These were extended to a number of different cosmological models by
Moscardini et al. (1998).  In particular we are interested in the
transient model of Matarrese et al., so called because the model does
not require a normalization at $z=0$.  In this model, one assumes that
all objects exceeding a given mass cut off can be observed at any
given redshift.  The bias (which we call model 2) then has the form
\begin{equation}
b(z)=1-1/\deltac+[b(0)-(1-1/\deltac)]G(\Om,\lo,z)^\beta
\label{biasmodel_2}
\end{equation}
where $\deltac$ is the critical linear overdensity for spherical
collapse.  For an EdS cosmology $\deltac=1.686$ for all redshifts,
however it only varies away from this value by a few per cent for the
other cosmologies considered here \cite{l92}.  Matarrese et al. find
the values of $b(0)$ and $\beta$ by fitting to their Press-Schechter
based models.  These parameters depend on the minimum halo mass
$\mmin$ considered.  We compare this model of biasing to QSO
clustering is an EdS universe in Fig. \ref{r0evol}a for minimum halo
masses of $\mmin=10^{11}$, $10^{12}$ and $10^{13}\msun$.  In this
cosmology the data are approximately consistent with a minimum halo
mass of $10^{12}\msun$ (although the model is still too steep), while
the normalization is too low (\cobe  normalization) for lower mass
haloes, and the evolution is too steep for higher mass haloes.  In the
$\Lambda$ cosmology, we compare the COBE normalized $\Lambda$CDM model
of Moscardini et al. to our data.  We note that this model is, in
fact, for $\Om=0.4$, $\lo=0.6$.  However,  given the model and data
uncertainties these are adequate to make a general comparison to the
2QZ clustering evolution in the $\Lambda$ cosmology.  In this case we
find that the data are more consistent with (although slightly above)
a model with $\mmin\simeq10^{13}\msun$.

Although these models appear to adequately describe the clustering
evolution of QSOs, it is not at all clear what the physical
justification for this is.  The models of Matarrese et al. assume that
at each redshift QSOs inhabit the same mass haloes; this need not
necessarily be the case.  For example, Percival \& Miller (1999)
compare the evolution of bright QSOs, $-25.4>M_{\rm B}>-27.9$, to the
dark matter halo formation rate in a number of cosmologies.  They find
that for an EdS universe, with a CDM-type power spectrum of shape
parameter $\Gamma_{\rm eff}=0.25$ which is cluster abundance
normalized, the evolution of bright QSOs is best fit by haloes of mass
$\sim10^{10.6}\msun$.  Our $\Lambda$ cosmology increases the mass to
$\sim10^{11.8}\msun$.  These masses are $\sim10\times$ smaller than
those required to fit the 2QZ QSO clustering according to the models
of Matarrese et al.  This serves to demonstrate that we should be wary
of over interpreting fits to models which do not contain a physical
description of QSO formation.  For example, it is possible that QSO
clustering is a function of luminosity, a point which has not been
discussed in this paper, but will be investigated in future work.

\subsubsection{An empirical biasing description}

Lastly we fit a purely empirical biasing model to the data.  For this
model we use a generalization of Eqs. \ref{biasmodel_1} and
\ref{biasmodel_2} which is
\begin{equation}
b(z)=1+(b(0)-1)G(\Om,\lo,z)^\beta,
\label{biasmodel_3}
\end{equation}
where $b(0)$ and $\beta$ are left free to be determined by fitting to
the data.  We call this form of bias evolution model 3.  The
normalization of the mass density field is set by  either cluster or
{\it COBE} normalization as in model 1.  The dot-dash lines in
Fig.\ref{r0evol} show the best fit empirical model for each of our
assumed cosmologies.  In the EdS case we find
$b(0)=1.45^{+0.21}_{-0.16}$ and $\beta=1.68^{+0.44}_{-0.40}$
($b(0)=1.28^{+0.16}_{-0.11}$ and $\beta=1.89^{+0.49}_{-0.46}$) for
cluster (COBE) normalization.  As we might expect the $\Lambda$
cosmology has a larger $\beta$ with the best fit parameters being
$b(0)=1.20^{+0.06}_{-0.02}$ and $\beta=2.75^{+0.65}_{-0.57}$.  The
relatively high normalization in this cosmology and the slow rate of
mass clustering evolution means that a large value of $\beta$ is
required to fit the data.

\section{conclusions}\label{section_conclusions}

The preliminary release dataset of the 2QZ contains 10681 QSOs.  It is
already a factor of $\sim25$ larger than previous QSO surveys to this
depth ($\bj\leq20.85$).  When completed the full sample will contain
$\sim25000$ QSOs.  The current data set already allows us to measure
the clustering of QSOs to un-precedented accuracy.  In particular we
find:

1) QSO clustering integrated over the redshift interval $0.3<z\leq2.9$
is well fit by a power law on scales $\sim1-35\Mpc$.  In an
Einstein-de Sitter universe the best fit power law has
$s_0=3.99^{+0.28}_{-0.34}\Mpc$ and $\gamma=1.58^{+0.10}_{-0.09}$.
Introducing a cosmological constant increases the distances between
QSOs, so that the scale length of clustering increases also.  The
power law then extends to $\sim60\Mpc$ and is best fit by
$s_0=5.69^{+0.42}_{-0.50}\Mpc$ and $\gamma=1.56^{+0.10}_{-0.09}$.
These results are remarkably similar to the clustering of normal
galaxies locally ($z\simeq0.05$).  

2) We compare the clustering of 2QZ QSOs to the $\Lambda$CDM model and
find that the shapes of model and data are consistent.  A comparison
to a family of CDM models with different shape parameters,
$\Gamma_{\rm eff}$, finds that $\Gamma_{\rm eff}=0.2$ to 0.4 provides
an acceptable fit in the EdS cosmology.  In the $\Lambda$ cosmology
only $\Gamma_{\rm eff}=0.1$ or 0.2 provide an acceptable fits due to
the movement of structure to larger scales.   This suggests a test for
cosmological parameters using the linear break in the correlation
function which will be possible using the completed 2QZ data set.

3) We measure the clustering amplitude of QSOs as a function of
redshift, parameterized by $s_0$ assuming a fixed power law slope.  In
an Einstein-de Sitter universe we find that QSO clustering is constant
in comoving coordinates over the entire redshift range we probe.  In a
$\Lambda$ dominated universe we find that clustering appears to
increase (although constant clustering is not excluded) with
increasing redshift.  For both EdS and $\Lambda$ cosmologies a model
in which QSOs follow the same evolution as linear theory gravitational
clustering (or have a bias which is constant as a function of
redshift) is rejected at the $>99$ per cent level.  If the constant
clustering is extrapolated to $z\simeq3$ it comfortably overlaps the
clustering amplitude found for Lyman-break galaxies
\cite{adelberger98}.

4) We compare simple redshift dependent bias models to the measured
clustering evolution.  We first use a model in which QSOs are long
lived (on cosmological time scales), so that their clustering simply
evolves according to their motion in the gravitational potential.
This is consistent with 2QZ clustering evolution in an EdS case, and
predicts $\sigma_8(z=0)\simeq1$ for QSOs, which is consistent with
galaxy clustering.  The long lived  model is not able to reproduce the
increase in clustering seen in the $\Lambda$ cosmology and is
marginally rejected at 88 per cent confidence.  More complex models of
QSO bias based on the Press-Schechter formalism, have been developed
by a number of authors.  We use the models of Matarrese et al. (1997)
and Moscardini et al (1998) to make comparisons to the evolution of
the 2QZ data set.  These models  adequately describe the 2QZ
clustering evolution when the minimum halo mass considered is
$\mmin\sim10^{12}\msun$ (EdS) or $\mmin\sim10^{13}\msun$ ($\Lambda$).
However, without a convincing model of QSO formation, the
interpretation of the comparison to these models of clustering
evolution is questionable.  We lastly derive a fit to an empirical
biasing model based on power law evolution of bias.

The large volumes sampled by QSO surveys allow structure to be
investigated on the scales where growth is governed by linear theory.
Thus, meaningful measurements of large-scale structure, that are
easily related to the underlying cosmology, can be made irrespective
of the relative bias of QSOs.  QSOs therefore play an crucial role in
linking low-redshift/small-scale galaxy clustering measurements to the
fluctuations in the density field at high redshift seen in the cosmic
microwave background.  The completed 2QZ survey, without the current
varying observational coverage, will allow detailed measurements of
structure on a range of scales from $\sim1$ to $1000\Mpc$.

\section*{acknowledgments}

This paper was prepared using facilities of the Anglo-Australian
Observatory and the Starlink node at the Imperial College of Science
Technology and Medicine.  The 2QZ is based on observations made with
the Anglo-Australian Telescope and the UK Schmidt Telescope.  NSL and
FH are supported by PPARC Studentships.


\begin{thebibliography}{}

\bibitem[Adelberger et al. 1998]{adelberger98} Adelberger K.L.,
 Steidel C.C., Giavalisco M., Dickinson M., Pettini M., Kellogg M.,
 1998, ApJ, 505, 18

\bibitem[Alcock \& Paczynski 1979]{ap79} Alcock C., Paczynski B.,
1979, Nature, 281, 358

\bibitem[Andreani \& Cristiani 1992]{ac92} Andreani P., Cristiani S.,
1992, ApJ, 398, L13

\bibitem[Bailey \& Glazebrook 1999]{2dfman} Bailey J., Glazebrook K.,
1999, 2dF User Manual, Anglo-Australian Observatory

\bibitem[Ballinger et al. 1996]{bph96} Ballinger W.E., Peacock J.A.,
Heavens A.F., 1996, MNRAS, 282, 877

\bibitem[Bennett et al. 1996]{cobe96} Bennett C.L. et al., 1996, ApJ,
464, L1

\bibitem[Boyle et al. 1990]{bfsp90} Boyle B.J., Fong R., Shanks T.,
Peterson B.A., 1990, MNRAS, 243, 1

\bibitem[Boyle et al. 1995]{phot1} Boyle B.J., Shanks T.,
Croom S.M., 1995, MNRAS, 276, 33

\bibitem[Boyle et al. 2000]{paper1} Boyle B.J., Shanks T.,
Croom S.M., Smith R.J., Miller L., Loaring N., Heymans C., 2000,
MNRAS, 317, 1014

\bibitem[Carlberg et al. 1999]{cnoc2} Carlberg R.G. et al., 1999,
Phil. Trans. Royal Society, 357, 167

\bibitem[Carroll et al. 1992]{cpt92} Carroll S.M., Press W.H., Turner
E.L., 1992, ARA\&A, 30 , 499

\bibitem[Colberg et al. 1998]{hubvol98} Colberg J.M., 1998, in Wide
Field Surveys in Cosmology, 14th IAP meeting, Editions Frontieres,
Paris, 247

\bibitem[Colless 1999]{2dfgrs} Colless M., 1999, in Morganti R., Couch
 W.J., eds., Proc. ESO/Australia Workshop, Looking Deep in the
 Southern Sky. Springer-Verlag, p.9

\bibitem[Croom 1997]{croom} Croom S.M., 1997, Ph.D. Thesis, University
of Durham

\bibitem[Croom \& Shanks 1996]{cs96} Croom S.M., Shanks T., 1996,
MNRAS, 281, 893

\bibitem[Croom et al. 1999]{phot2} Croom S.M., Ratcliffe
A., Parker Q.A., Shanks T., Boyle B.J., Smith R.J., 1999, MNRAS, 306, 592

\bibitem[Efstathiou \& Rees 1988]{er88} Efstathiou G., Rees M.J.,
1988, MNRAS, 230, 5P

\bibitem[Eke et al. 1996]{ecf96} Eke V.R., Cole S., Frenk C.S., 1996,
MNRAS, 282, 263

\bibitem[Fry 1996]{f96} Fry N.J., 1996, ApJ, 461, L65

\bibitem[Gehrels 1986]{g86} Gehrels N., 1986, ApJ, 303, 336

\bibitem[Hamilton et al. 1991]{hklm91} Hamilton A.J.S., Kumar P., Lu
E., Matthews A., 1991, ApJ, 374, L1

\bibitem[Iovino \& Shaver 1988]{is88} Iovino A., Shaver P.A., 1988,
ApJ, 330, L13

\bibitem[Jain et al. 1995]{jmw95} Jain B., Mo H.J., White S.D.M.,
1995, MNRAS, 276, L25

\bibitem[Kaiser 1987]{kaiser87} Kaiser N., 1987, MNRAS, 227, 1

\bibitem[La Franca et al. 1998]{lac98} La Franca F., Andreani P.,
Cristiani, S., 1998, ApJ, 497, 529

\bibitem[Landy \& Szalay 1993]{ls93} Landy, S.D., Szalay, A.S., 1993,
ApJ, 412, 64

\bibitem[Le Fevre et al. 1996]{cfrs96} Le Fevre O., Hudon D., Lilly
S.J., Crampton D., Hammer F., Tresse L., 1996, ApJ, 461, 534

\bibitem[Lilje 1992]{l92} Lilje P.B., 1992, ApJ, 386, L33

\bibitem[Loveday et al. 1995]{lmep95} Loveday J., Maddox S.J.,
Efstathiou G., Peterson B.A., 1995, ApJ, 442, 457

\bibitem[Maddox et al. 1990]{mesl90} Maddox S.J., Efstathiou G.,
Sutherland W.J., Loveday J., 1990, MNRAS, 242, 43P

\bibitem[Magorrian et al 1998]{mag98} Magorrian J. et al., 1998, AJ,
115, 2285

\bibitem[Mann et al. 1998]{mph98} Mann R.G., Peacock J.A., Heavens
A.F., 1998, MNRAS, 293 209

\bibitem[Matarrese et al. 1997]{mclm97} Matarrese S., Coles P.,
Lucchin F., Moscardini L., 1997, MNRAS, 286, 115

\bibitem[Moscardini et al. 1998]{mclm98} Moscardini L., Coles P.,
Lucchin F., Matarrese S., 1998, MNRAS, 299, 95

\bibitem[Mo et al. 1992]{mjb92} Mo H.J., Jing Y.P., Borner G., 1992,
ApJ, 392, 452

\bibitem[Mo \& Fang 1993]{mf93} Mo H.J., Fang L.Z., 1993, ApJ, 410, 493

\bibitem[Mo \& White 1996]{mw96} Mo H.J., White S.D.M., 1996, MNRAS,
282, 347

\bibitem[Osmer 1981]{o81} Osmer P.S., 1981, ApJ, 247, 762

\bibitem[Peacock 1997]{p97} Peacock J.A., 1997, MNRAS, 284, 885

\bibitem[Peacock \& Dodds 1996]{pd96} Peacock J.A., Dodds S.J., 1996,
MNRAS, 280, L19

\bibitem[Percival \& Miller 1999]{pm99} Pervical W., Miller L., 1999,
MNRAS, 309, 823

\bibitem[Phillipps 1994]{p94} Phillipps S., 1994, MNRAS, 269, 1007

\bibitem[Postman et al. 1998]{plso98} Postman M., Lauer T.R., Szapudi
I., Oegerle W., 1998, ApJ, 506 33

\bibitem[Press \& Schechter 1974]{ps74} Press W.H., Schechter P.,
1974, ApJ, 187, 425

\bibitem[Ratcliffe et al. 1998]{rspf98} Ratcliffe A., Shanks T., Parker Q.A.,
Fong R., 1998, MNRAS, 296, 173

\bibitem[Schade et al 2000]{sbl00} Schade D., Boyle B.J., Letawsky M.,
2000, MNRAS in press

\bibitem[Schlegel et al. 1998]{schlegel98} Schlegel D.J., Finkbeiner
D.P., Davis M., 1998, ApJ, 500, 525

\bibitem[Shanks et al. 1987]{sfbp87} Shanks T., Fong R., Boyle B.J.,
Peterson B.A., 1987, MNRAS, 227, 739

\bibitem[Shanks \& Boyle 1994]{sb94} Shanks T., Boyle B.J., 1994,
MNRAS, 271, 753

\bibitem[Shaver 1984]{s84} Shaver P.A., 1984, A\&A, 136, L9

\bibitem[Smith 1998]{smith} Smith R.J., 1998, Ph.D. Thesis, University
of Cambridge

\bibitem[Smith et al. 2001]{catpaper} Smith R.J., Croom S.M., Boyle
B.J., Shanks T., Miller L., Loaring N.A., 2001, MNRAS, submitted.

\bibitem[Steidel et al. 1998]{steidel98} Steidel C.C., Adelberger
 K.L., Dickinson M., Giavalisco M., Pettini M., Kellogg M., 1998, ApJ,
 492, 428 

\bibitem[Sugiyama 1995]{sug95} Sugiyama N., 1995, ApJS, 100, 281

\bibitem[Taylor et al. 1997]{taylor97} Taylor K., Cannon R.D., Watson
F.G., 1997, in Ardeberg A.L., Ed., Proc. SPIE conf. Optical Telescopes
of Today and Tomorrow, 2871, 145

\bibitem[Tucker et al. 1997]{tucker97} Tucker D.L. et al., 1997,
MNRAS, 285, L5

\bibitem[Weinberg 1972]{weinberg72} Weinberg S., 1972, Gravitation and
cosmology, Wiley

\end{thebibliography}
\end{document}